\documentclass[pra,aps,%
 twocolumn,
superscriptaddress,
nofootinbib,
longbibliography]{revtex4-2}
\usepackage{makecell} 
\usepackage{braket}
\usepackage[table,dvipsnames]{xcolor}

\usepackage{graphicx}
\usepackage{amsfonts}
\usepackage{amssymb}
\usepackage{booktabs}
\usepackage{amsmath}
\usepackage{amsthm}

\usepackage{booktabs}
\usepackage[colorlinks=true,linkcolor=teal,citecolor=teal,urlcolor=teal,plainpages=false,pdfpagelabels]{hyperref}
\usepackage{color}

\usepackage{mathtools}
\usepackage{bbm}
\usepackage{textcomp}

\usepackage[caption=false]{subfig}

\usepackage{array}

\usepackage{enumitem}
\usepackage{verbatim}

\usepackage[utf8]{inputenc}
\usepackage[T1]{fontenc}
\usepackage{etoolbox}

\usepackage{newtxtext}
\usepackage{newtxmath}
\usepackage[normalem]{ulem}

\usepackage[utf8]{inputenc}
\usepackage[T1]{fontenc}

\makeatletter
\g@addto@macro\bfseries{\boldmath}
\makeatother

\newcommand{\Tr}{\mathrm{Tr}}

\theoremstyle{definition}

\renewcommand{\qedsymbol}{$\blacksquare$}

\renewcommand{\qedsymbol}{\unskip\nobreak\quad\qedsymbol}
\renewcommand{\qedsymbol}{$\blacksquare$}

\DeclareMathOperator*{\argmax}{arg\,max}

\definecolor{dblue}{rgb}{0.0, 0.53, 0.74}

\begin{document}

\title{Feature-level analysis and adversarial transfer in rotationally equivariant \\ quantum machine learning}

\author{Maureen Krumt{\"u}nger}
\email{maureen.krumtuenger@student.unimelb.edu.au}
\affiliation{School of Physics, University of Melbourne, Parkville, VIC 3010, Australia}
 
\author{Martin Sevior}
\affiliation{School of Physics, University of Melbourne, Parkville, VIC 3010, Australia}
 
\author{Muhammad Usman}
\affiliation{School of Physics, University of Melbourne, Parkville, VIC 3010, Australia}
\affiliation{Data61, CSIRO, Clayton, 3168, Victoria, Australia}


\begin{abstract}
\noindent
Group-equivariant quantum models are designed to exploit symmetry and can improve trainability, but it remains unclear how symmetry constraints shape their adversarial robustness. We study this question through a feature-level analysis of equivariant quantum models in a transfer-attack setting. Under equivariance with an invariant readout, predictions depend only on the group-twirled input, which identifies the symmetry-invariant information accessible to the model together with a complementary uninformative subspace. Specializing this framework to a rotationally equivariant quantum model, we derive an explicit characterization of the accessible information in terms of rotation-invariant image statistics distributed across distinct symmetry sectors. Using targeted input transformations, we determine which of these statistics are actually relied upon for classification across several datasets. We find that equivariance alone does not guarantee transfer robustness: even within the restricted invariant feature space, the model can rely on brittle statistics, particularly ring-averaged intensities in the rotationally equivariant model, that remain vulnerable to classical transfer attacks. Guided by this analysis, we show that suppressing the symmetry sector associated with the brittle feature substantially improves robustness. These results establish a systematic mechanism to exploit symmetry-dependent features for adversarial robustness in future quantum machine learning models.
\end{abstract}

\maketitle

\section{Introduction}

Many quantum machine learning (QML) tasks have geometric structure: labels may be invariant under transformations such as rotations, permutations, or translations. Geometric QML uses group-equivariant architectures to improve inductive bias and, in several constructions, guarantee trainability~\cite{Larocco_2022, Meyer2023, Nguyen2024, Schatzki_2024, West_2024, West_2023_reflect}. Much of this progress has relied on circuit-level analyses using Lie-algebraic and representation-theoretic tools to study expressivity and gradient scaling. Yet, even for well-understood architectures, comparatively little attention has been paid to which input features are actually accessible to the model and ultimately drive its predictions. Such a mechanistic understanding is valuable not only for interpretability and ultimately explainability of QML models, but also for analyzing properties beyond trainability, such as adversarial robustness.

Here, adversarial robustness refers to a model’s resistance to small, carefully crafted input perturbations that induce misclassifications~\cite{yuan2018, ibitoye2023}. It represents a key requirement for the deployment of machine learning systems in safety-critical settings such as self-driving vehicles and autonomous military systems. At the same time, the existence of adversarial examples is itself an intriguing phenomenon that has been studied extensively in classical machine learning~\cite{goodfellow2015, tsipras2019,ilyas2019, demontis2019, madry2019}. Particularly relevant to this work is the feature-based viewpoint, according to which the features used by a classifier can differ fundamentally in their robustness~\cite{ilyas2019, tsipras2019}. In this picture, adversarial vulnerability arises when predictions rely on brittle, non-robust features, whereas robustness may emerge when a model becomes invariant to, or suppresses reliance on, such features~\cite{tsipras2019}.

Adversarial robustness in QML has also received significant attention in recent years~\cite{usman2026}. Initial works suggested that quantum models are also vulnerable to adversarial examples~\cite{Lu_2020,Gong_2021}.
However, recent studies have shown that quantum models can exhibit increased resilience to transfer attacks crafted on classical surrogate models~\cite{West_2023, West_2023to, Dowling_2026}. One possible explanation is that quantum and classical learners rely on different predictive features of the same data, as evidenced by the qualitatively different perturbation patterns observed for attacks optimized on classical versus quantum models~\cite{West_2023}. This naturally raises the question of how equivariance shapes the predictive features accessible to a quantum model and, in turn, affects adversarial transfer robustness.

In this work, we provide a feature-level characterization of the information accessible to an equivariant quantum model and examine how this relates to adversarial transfer robustness. For this purpose, we focus on the rotationally equivariant quantum model introduced in Ref.~\cite{West_2024}, which offers a particularly transparent setting because of its symmetry structure and provable trainability. While Ref.~\cite{West_2024} introduced this architecture and established its trainability, our contribution is different in scope: we derive a twirling-based characterization of the input information the model can access, thereby obtaining a much clearer understanding of how the model works. We show that the model only accesses rotationally invariant image statistics that decompose across its symmetry channels. To probe the role of these statistics, we introduce controlled input transformations that selectively preserve or disrupt specific invariant statistics. These transformations allow us to validate our characterization numerically and to identify which accessible features are most relevant for classification across datasets. Although the feature-level interpretation we develop is specific to this architecture and encoding, the general twirling-based framework applies more broadly to strictly equivariant quantum models.

Our characterization also enables us to address a question that, to our knowledge, has not previously been studied for symmetry-equivariant quantum models: their adversarial transfer robustness.
To this end, we evaluate transfer robustness against attacks crafted on classical surrogate models using the fast gradient sign method (FGSM)~\cite{goodfellow2015} and projected gradient descent (PGD)~\cite{madry2019}. Our results show that equivariance alone does not guarantee transfer robustness: even within the restricted space of rotationally invariant statistics, the model can still rely on brittle features. For the rotationally equivariant model studied here, these features are primarily associated with ring-averaged intensities, although their importance depends strongly on the dataset. Finally, we show that transfer robustness can be improved substantially by suppressing the symmetry channel associated with these ring-averaged intensities. Overall, our study provides a concrete understanding of adversarial attack transferability and shows how feature-level analysis can guide robustness improvements in symmetry-equivariant quantum models.

The remainder of this paper is organized as follows. Section~\ref{sec:background} reviews the rotationally equivariant quantum model of Ref.~\cite{West_2024} and the adversarial attacks considered. In Section~\ref{sec:methods}, we develop a twirling-based characterization of the input information accessible to the model, interpret it at the pixel level, and introduce diagnostic input transformations. Section~\ref{sec:numerical_setup} describes the numerical setup. Section~\ref{sec:results} presents the adversarial-robustness results under transfer attacks crafted on classical surrogate models, and relates these empirical findings to the feature-level analysis. We conclude with a discussion of the implications of our study in Section~\ref{sec:discussion}.

\section{Preliminaries}\label{sec:background}
This section summarizes key concepts of group-equivariant quantum models, especially focusing on the rotationally equivariant model of Ref.~\cite{West_2024} and the adversarial attacks considered for our robustness evaluation. More comprehensive treatments of the broader framework of geometric QML are available in Refs.~\cite{Meyer2023, Larocco_2022, Nguyen2024}. 

\subsection{Group-equivariant quantum models for supervised learning}\label{sec:bg_re}

In the supervised learning setting, a dataset $\mathcal{D}=\{(x_i,y_i)\}_{i=1}^N$ is given, and the goal is to assign each input $x\in\mathcal{X}$ to its associated label $y(x)\in\mathcal{Y}=\{1,\dots,n_{\text{classes}}\}$, among $n_{\text{classes}}$ classes. A quantum classifier for this task consists of an encoding unitary $\mathcal{E}(x)$ that prepares
$\ket{\psi(x)} := \mathcal{E}(x)\ket{0}$, 
a trainable variational unitary $U_{\boldsymbol{\theta}}$, and measurement operators $\{M_j\}$.
By measuring a collection of measurement operators 
$\{ M_j \}_{j=1}^{n_{\text{classes}}}$ associated with the distinct classes, 
the model's prediction $\hat{y}_\theta(x)$ is defined as the class whose 
measurement operator yields the largest expectation value:

\begin{equation}
\hat{y}_{\boldsymbol{\theta}}(x)
= \argmax_j \Tr\,\!\Big[M_j\, U_{\boldsymbol{\theta}} \rho(x) U_{\boldsymbol{\theta}}^\dagger \Big],
\end{equation}
where we use $\rho(x) := \ket{\psi(x)}\!\bra{\psi(x)}$.

Now suppose that the task is invariant under a group  $\mathcal{G}$ acting on $\mathcal{X}$, i.e.,
$y(x)=y(gx)$ for all $x\in\mathcal{X}$ and $g\in\mathcal{G}$.
A $\mathcal{G}$-equivariant quantum model is designed such that its predictions are invariant under the action of the symmetry group, i.e.,
\[
\hat{y}_{\boldsymbol{\theta}}(x)=\hat{y}_{\boldsymbol{\theta}}(g x),
\qquad \forall\, x\in\mathcal{X},\; g\in\mathcal{G},\; \boldsymbol{\theta}\in\Theta.
\]
A sufficient condition for invariance is (see Ref.~\cite{West_2024} for more details):
\begin{align}
    [\,R(g),\, &U_{\boldsymbol{\theta}}^{\dagger} M_j U_{\boldsymbol{\theta}}\,]
    = 0,\label{eqqui_condition}\\
    &\qquad
    \forall\, g\in\mathcal{G},\;
    \boldsymbol{\theta}\in\Theta,\;
    j\in\{1,\ldots,n_{\mathrm{classes}}\}\notag,
\end{align}
where $R(g)$ is a unitary representation of the group action $g\in\mathcal{G}$ on $\mathcal{H}$, induced by the selected encoding map $\mathcal{E}$. Multiple strategies for designing quantum models that satisfy the invariance condition~\eqref{eqqui_condition} have been proposed, for example, in Refs.~\cite{Larocco_2022, Nguyen2024}.

\subsubsection{The rotationally equivariant quantum model.}\label{sec:rotequiv}

\begin{figure}
    \centering
    \includegraphics[width=\linewidth]{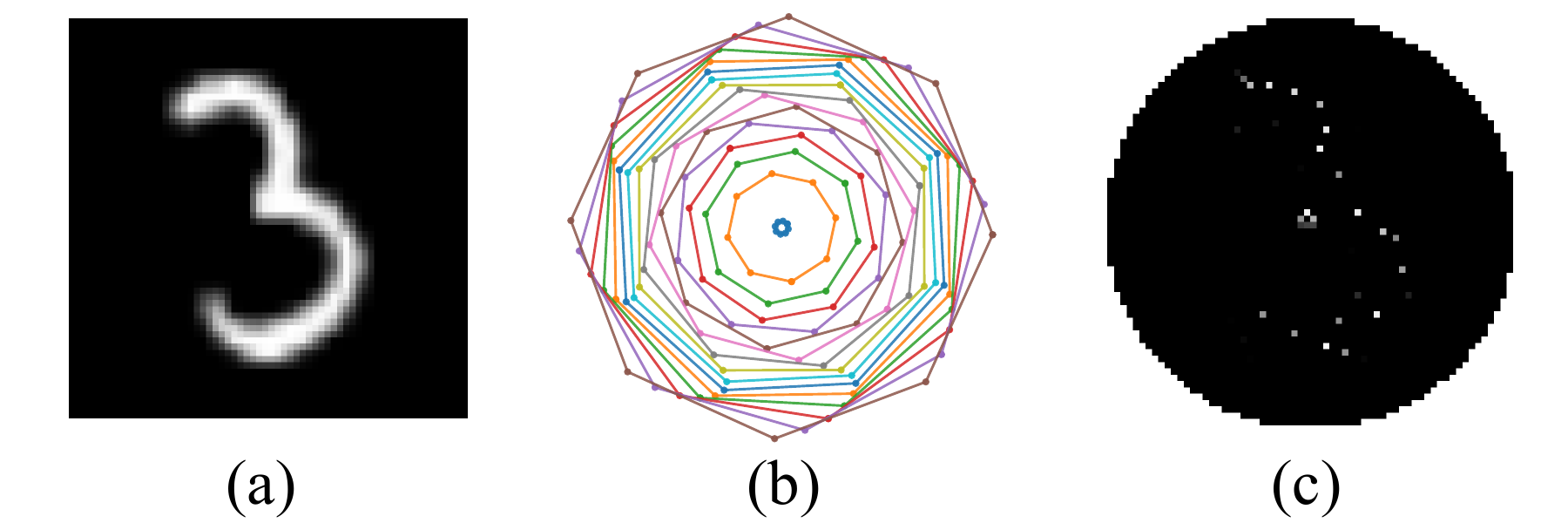}
    \caption{Illustration of the encoding scheme for a representative sample from the MNIST dataset (see Sec.~\ref{sec:numerical_setup} for details on the dataset).
    (a) original image.
    (b) sampling grid induced by the encoding for $n_{\mathrm{orb}}=3$ and $n_{\mathrm{rad}}=4$, consisting of $N_r=16$ polygonal rings with $N_\phi=8$ vertices each; each vertex corresponds to a sampled pixel value $x_{r,\phi}$.
    (c) corresponding encoded image obtained by placing the sampled values at the vertex locations (all other pixels set to a constant background).
    For clarity, we visualize a sparse setting; increasing $n_{\mathrm{rad}}$ and/or $n_{\mathrm{orb}}$ increases the sampling density and thus the effective resolution of the image. }
    \label{fig:encoding_scheme}
\end{figure}

Specifically, the authors of Ref.~\cite{West_2024} construct a rotationally equivariant quantum model by partitioning a $n$-qubit register into a radial register consisting of $n_{\mathrm{rad}}$-qubits and an orbital register consisting of $n_{\mathrm{orb}}$-qubits. Based on this decomposition, the encoding scheme samples the pixel values of an image $x$ at vertices obtained by constructing $N_{r}=2^{n_{\mathrm{rad}}}$ regular polygons (which we refer to as \emph{rings} in the following), each with $N_{\phi}=2^{n_{\mathrm{orb}}}$ vertices. An illustrative example of the encoding scheme is shown in Figure~\ref{fig:encoding_scheme}.

The resulting pixel values $x_{r,\phi}$, where
$r \in \{0,1,\ldots,2^{n_{\mathrm{rad}}}-1\}$ labels the rings and
$\varphi_\phi = 2\pi \phi / N_\phi$ with $\phi \in \{0,1,\ldots,N_\phi-1\}$ denotes the angular position on each ring, are then used to construct the quantum state 
\begin{align}
\ket{\psi(x)} = \frac{1}{\|x\|} \sum_{r,\phi} x_{r,\phi} \ket{r,\phi}.
\label{enc_state}
\end{align}

Based on this encoding, rotations of the image $x$ are represented by the action of the cyclic group $\mathbbm{Z}_{N_{\phi}}$ of order $N_{\phi}$, which acts through its regular representation $R(g)$ exclusively on the orbital register as a cyclic shift of the angular index~$\phi$:
\begin{align}
    R(g)\ket{\psi(x)}=\frac{1}{\|x\|}\sum_{r,\phi}x_{r,\phi}\ket{r,\phi+g\,(\text{mod}\, N_{\phi})}.
\end{align}

Next, the quantum Fourier transform (QFT) over the cyclic group $\mathbbm{Z}_{N_{\phi}}$ is applied to the orbital register, i.e. $(\mathbbm{I}_{\text{rad}}\otimes \text{QFT}^{\dagger}_{\text{orb}})\circ\mathcal{E}(x)$, thereby expressing the orbital register in the Fourier basis of the group action. Since $\mathbbm{Z}_{N_\phi}$ is abelian, all of its irreducible representations (irreps) are one-dimensional and labeled by $m\in{0,\dots,N_\phi-1}$. Each Fourier mode $m$ therefore corresponds to one irrep sector.
As a consequence, constructing a variational layer that commutes with the group action is straightforward: any gate acting diagonally on the orbital register in the Fourier basis automatically lies in the commutant of the representation. Similarly, the measurement operator is chosen as
$M_j := Z_j \quad (j=1,\dots,n_{\text{classes}})
$, with $n_{\text{classes}}\leq n_{\text{rad}}$, which also commutes with the group action. Appendix~\ref{app:rot_equiv_model_architecture} provides a schematic overview of the full architecture.

\subsection{Adversarial attacks}\label{sec:adv_attacks_theory}

Adversarial attacks are deliberately crafted input perturbations intended to mislead a machine learning model. In this work, we consider two widely used white-box attacks, which are applied only to classical surrogate models and evaluated in a transfer-only setting on the target quantum model.

\paragraph{Threat model.}
Given a clean input $x$, the adversary aims to construct a perturbed example
$x'=x+{\delta}$ with $\|{\delta}\|_{\infty}\le \varepsilon$
that induces misclassification of the target model $f$ at test time.
The adversary has \emph{no direct access} to $f$: it does not observe the model parameters, loss, or gradients, and it has no query access to $f$.
Instead, the adversary trains a \emph{classical surrogate} $\hat f$ on data drawn from the same training distribution and crafts perturbations by applying FGSM and PGD attacks to $\hat f$, which are then transferred to $f$.

\paragraph{Fast Gradient Sign Method (FGSM).} FGSM is a white-box attack that perturbs the input in the sign direction of the loss gradient, scaled by the attack budget $\varepsilon$.
The FGSM attack on a clean image $x$ is defined as \cite{goodfellow2015}
\begin{align}
        x_{\text{adv}}=x+\varepsilon\cdot \text{sign}\big(\nabla_x L(\boldsymbol{\theta}^*,x,y)\big),
\end{align}
where $\varepsilon$ denotes the $\ell_\infty$ attack budget, $L$ is the loss function of the model and $\boldsymbol{\theta}^*$ are the optimal parameters of the model.

\paragraph{Projected Gradient Descent (PGD).}
PGD is a multi-step variant of the FGSM attack \cite{madry2019}. 
Starting from an initial perturbation $x^0$, it iteratively updates the adversarial example as
\begin{align}
x^{t+1}
= \Pi_{\mathcal{B}_\varepsilon(x)}\!\Big(x^{t} + \alpha \,\mathrm{sign}\big(\nabla_x L(\boldsymbol{\theta}^*, x^{t}, y)\big)\Big),
\end{align}
where $\Pi_{\mathcal{B}_\varepsilon(x)}$ denotes the projection onto the $\ell_\infty$-ball 
$\mathcal{B}_\varepsilon(x) = \{x' : \|x' - x\|_\infty \le \varepsilon\}$ around the original input $x$, 
$\alpha$ is the step size, and $L$ is the loss function of the model. PGD is a strong first-order adversary: Ref.~\cite{madry2019} showed that, under certain conditions, PGD is a (near-optimal) universal first-order adversary in the sense that other polynomially-bounded first-order methods are unlikely to find consistently stronger perturbations.

\section{Theory and Methods}\label{sec:methods}

In this section, we characterize which input information is accessible to a model that enforces group equivariance, and we identify complementary directions in input space that are provably uninformative. Because any pixel-space interpretation necessarily depends on both the chosen encoding and the acting group, we carry out this analysis explicitly for the rotationally equivariant quantum model of Ref.~\cite{West_2024}.

\subsection{Information accessible to group-equivariant quantum models}
\label{sec:group_twirl}

Let $\mathcal{G}$ be a finite group with unitary representation $R(g)$ acting on the input Hilbert space. 
Assume (i) $U_{\boldsymbol{\theta}}$ is $\mathcal{G}$-equivariant, $U_{\boldsymbol{\theta}}R(g)=R(g)U_{\boldsymbol{\theta}}$, and (ii) the readout observable $M$ is $\mathcal{G}$-invariant, $R(g)^\dagger M R(g)=M$ for all $g\in \mathcal{G}$. 
Defining the twirling channel~\cite{Ragone_2023}
\begin{equation}
\mathcal T_{\mathcal{G}}(\rho):=\frac{1}{|\mathcal{G}|}\sum_{g\in \mathcal{G}} R(g)\rho R(g)^\dagger ,
\end{equation}
the prediction of the $\mathcal{G}$-equivariant quantum model depends on $\rho$ only through its $\mathcal{G}$-twirl:
\begin{equation}
f_{\boldsymbol{\theta}}(\rho)=f_{\boldsymbol{\theta}}\!\big(\mathcal T_{\mathcal{G}}(\rho)\big),
\label{eq:twirl_identity}
\end{equation}
where $f_{\boldsymbol{\theta}}(\rho)=\Tr[MU_{\boldsymbol{\theta}}\rho U^\dagger_{\boldsymbol{\theta}}]$.
This follows by averaging over $g$ and using linearity and cyclicity of the trace together with equivariance of $U_{\boldsymbol{\theta}}$ and invariance of $M$; see Appendix~\ref{group_twirl_to_input} for more details. 
Hence, the model cannot distinguish $\rho$ from its twirled counterpart $\mathcal T_{\mathcal{G}}(\rho)$, and any perturbation $\Delta$ at the state level satisfying $\mathcal T_{\mathcal G}(\Delta)=0$ lies in a provably uninformative subspace.

\subsubsection{Specializing to the rotationally equivariant model}
For the rotationally equivariant model of Ref.~\cite{West_2024}, described in Sec.~\ref{sec:bg_re}, the group $\mathcal{G}=\mathbbm{Z}_{N_\phi}$ with $N_{\phi}=2^{n_\text{orb}}$ acts as discrete rotations (cyclic shifts) on the orbital register. The goal of this section is to connect the Fourier-space description, in which the symmetry is naturally expressed, to the pixel-space description, in which the retained input information remains directly interpretable. A detailed derivation of this section is provided in Appendix~\ref{group_twirl_to_input}.

In the Fourier basis, the twirled density matrix takes the form
\begin{align}
\mathcal T_{\mathbb Z_{N_\phi}}(\rho)
=\frac{1}{\| x\|^2}\sum_{r,r'}\sum_{m=0}^{N_\phi-1}
\tilde x_{r,m}\tilde x^{*}_{r',m}\,\ket{r,m}\!\bra{r',m},
\label{eq:twirl_outcome_fourier}\end{align}
where $m\in\{0,\dots,N_\phi-1\}$ labels the irreducible representations of $\mathbb Z_{N_\phi}$ and the Fourier coefficients are defined as
\begin{equation}
\tilde x_{r,m}=\frac{1}{\sqrt{N_\phi}}\sum_{\phi=0}^{N_\phi-1} x_{r,\phi}\,\omega^{-m\phi},
\qquad \omega:=e^{2\pi i/N_\phi}.
\end{equation}

In particular, the trivial $m=0$ sector captures the ring-averaged intensities. Indeed,
\begin{equation}
\tilde x_{r,0}
=\frac{1}{\sqrt{N_\phi}}\sum_{\phi=0}^{N_\phi-1} x_{r,\phi}
=\sqrt{N_\phi}\,\bar x_r,
\end{equation}
and therefore
\begin{equation}
\tilde x_{r,0}\tilde x_{r',0}
= N_\phi\,\bar x_r\,\bar x_{r'},
\end{equation}
where we used that the pixel-values $x_{r,\phi}$ are real-valued.
Higher modes $m\neq 0$ encode structured intensity variations around each ring beyond the mean.

Equivalently, expressing the twirled state in the computational basis $\{\ket{r,\phi}\}_{r,\phi}$ of the radial and orbital registers yields
\begin{align}
\mathcal T_{\mathbb Z_{N_\phi}}(\rho)
&=\frac{1}{N_\phi\,\|x\|^2}\sum_{r,r'}\sum_{\phi,\Delta\phi}
\Bigg(\sum_{\alpha=0}^{N_\phi-1}x_{r,\alpha}\,x^{*}_{r',\alpha-\Delta\phi}\Bigg)\notag\\
&\qquad\qquad\qquad\qquad\qquad
\times\ket{r,\phi}\!\bra{r',\phi-\Delta\phi},
\label{eq:twirl_pixel_main}
\end{align}
where we set $\Delta\phi:=\phi-\phi'$ ($\bmod\,N_\phi$), so that $\phi'=\phi-\Delta\phi$.

Equation~\eqref{eq:twirl_pixel_main} makes explicit which input information remains accessible to the model under $\mathbb Z_{N_\phi}$-equivariance with an invariant readout: any dependence on the \emph{absolute} angle $\phi$ is averaged out, and the remaining coefficients depend only on the relative offset $\Delta\phi$. Equivalently, the accessible information is governed by the circular correlations
\begin{equation}
C_{r,r'}(\Delta\phi):=\sum_{\alpha=0}^{N_\phi-1} x_{r,\alpha}\,x^{*}_{r',\alpha-\Delta\phi},
\label{cross_ring}
\end{equation}
which are invariant under global rotations $x_{r,\phi}\mapsto x_{r,(\phi+g)\bmod N_\phi}$, since the sum runs over a full period.

\begin{figure}[t]
  \centering
    \includegraphics[width=\linewidth]{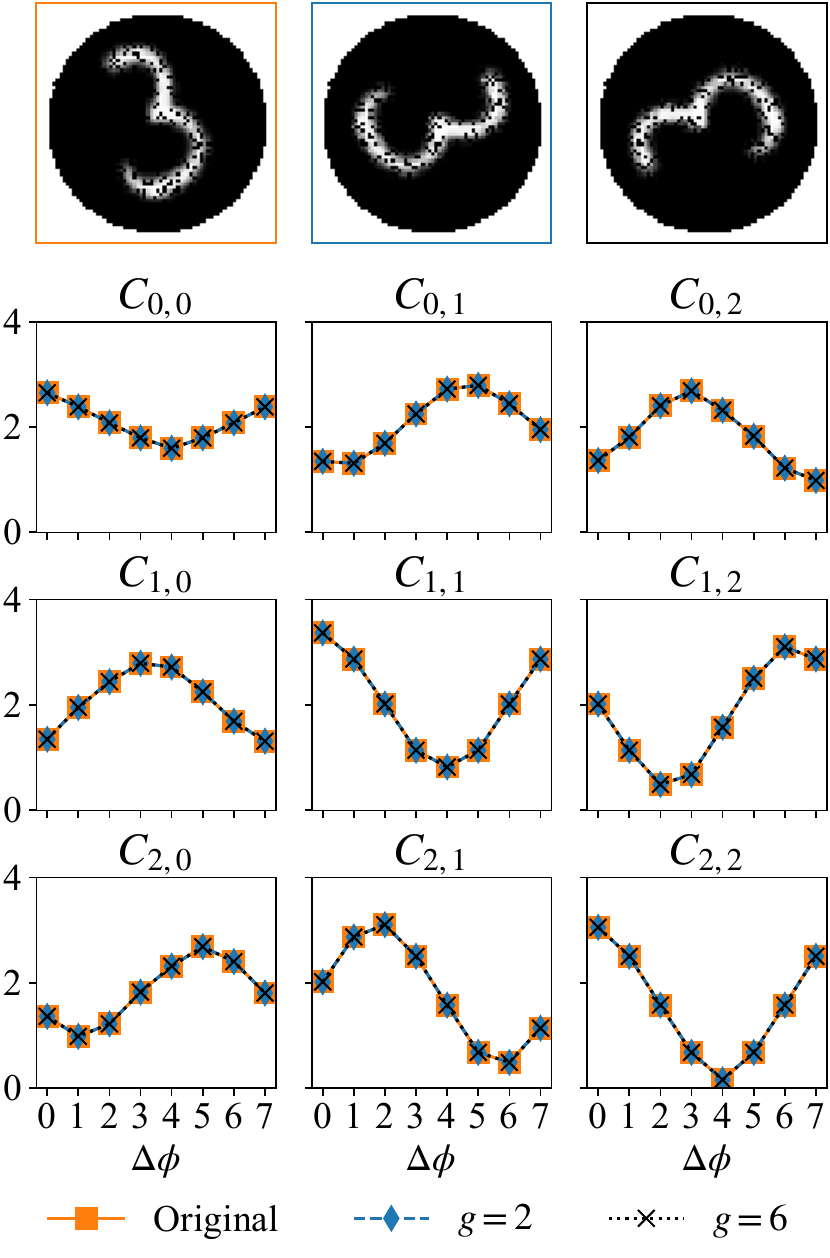}
    \caption{
    Illustration of circular correlations using a single sample from the MNIST dataset (see Sec.~\ref{sec:numerical_setup} for details on the MNIST dataset).
    The image is encoded with $N_r=512$ rings and $N_\phi=8$ angular samples per ring.
    Top: Original encoded input and two distinct rotations of the same sample.
    Bottom: The corresponding circular correlation functions $C_{r,r'}(\Delta\phi)$,
    computed via Eq.~\ref{cross_ring} for ring indices $r,r'\in\{0,1,2\}$
    for the original and rotated inputs.
    }
    \label{fig:interpretability}
\end{figure}

Figure~\ref{fig:interpretability} visualizes $C_{r,r'}(\Delta\phi)$ from Eq.~\eqref{cross_ring} on an MNIST sample (see Sec.~\ref{sec:numerical_setup} for details on the MNIST dataset). 
While the top row shows the original encoded input and two distinct rotations of the same sample, the bottom row reports $C_{r,r'}(\Delta\phi)$ for $r,r'\in\{0,1,2\}$. 
The correlations are unchanged by rotation, illustrating that—by design—the equivariant model can base its prediction only on rotation-invariant statistics. In contrast, a non-equivariant model can in principle leverage orientation-dependent features tied to an absolute reference angle $\phi$, which change under rotation.

\subsection{Probing the use of rotation-invariant statistics}

While Eq.~\ref{eq:twirl_pixel_main} characterizes which rotation-invariant statistics are, in principle, accessible to the equivariant model, it does not determine which of them are actually used after training or how they relate to robust and non-robust behavior. To address this, we introduce two complementary approaches to probe which accessible statistics are used by the trained model: (i) targeted input transformations and (ii) readout-level modifications.

\subsubsection{Input transformations.}
\label{sec:transformations_theory}

The characterization above yields a concrete prediction: a $\mathbbm{Z}_{N_\phi}$-equivariant model should be invariant under any input transformation that leaves the twirled representation $\mathcal T_{\mathbbm{Z}_{N_\phi}}(\rho)$ unchanged. Transformation~1 (T1) is designed to directly test this prediction. Further, to probe which rotation-invariant statistics are effectively used in practice, we additionally consider two complementary controls: a lossy mean-preserving transformation (T2) and a ring-mean removal transformation (T3). Visual examples for all transformations are shown in Figure~\ref{fig:transformed_figure}; full specifications are given in Appendix~\ref{app:transformations}.

In the following, we represent each image by its ring samples
\[
x_r=(x_{r,0},\cdots,x_{r,N_{\phi}-1})^T\in\mathbb{R}^{N_\phi},\quad  \forall  r\in\{0,\dots,N_r-1\}, 
\]
and define transformations ring-wise.

\paragraph{Transformation~1 (T1): orthogonal circulant scrambling.}
We construct a random orthogonal circulant map \(O(\lambda)\in\mathbb{R}^{N_\phi\times N_\phi}\) from a diagonal phase vector \(\lambda\), chosen such that \(O(\lambda)\) is real and orthogonal, and apply the same map to every ring,
\begin{equation}
x_r^{(1)} = O(\lambda)\,x_r \qquad \forall r\in\{0,\dots,N_r-1\}.
\end{equation}
By construction (see Appendix~\ref{app:transformations}), this leaves all rotation-invariant circular correlations unchanged and hence preserves $\mathcal T_{\mathbb Z_{N_\phi}}(\rho)$.

\paragraph{Transformation~2 (T2): ring-wise permutation.}
For each ring, we randomly sample a permutation matrix $P_r\in\mathbb{R}^{N_\phi\times N_\phi}$ to permute the angular order,
\begin{equation}
x_r^{(2)} = P_r x_r.
\end{equation}
This preserves ring averages but disrupts higher-order rotation-invariant correlation structure.

\paragraph{Transformation~3 (T3): ring-wise mean removal.}
We subtract the ring mean from each ring,
\begin{equation}
x_r^{(3)} = x_r - \bar x_r\,\mathbf{1},\qquad 
\bar x_r = \frac{1}{N_\phi}\sum_{\phi} x_{r,\phi}, 
\end{equation}
where $\mathbf{1}$ is an all-ones vector.
This eliminates the ring averages while retaining the correlation structure up to a $\Delta\phi$-independent offset (see Appendix~\ref{app:transformations}).

\begin{figure}[t]
    \centering
    \includegraphics[width=\linewidth]{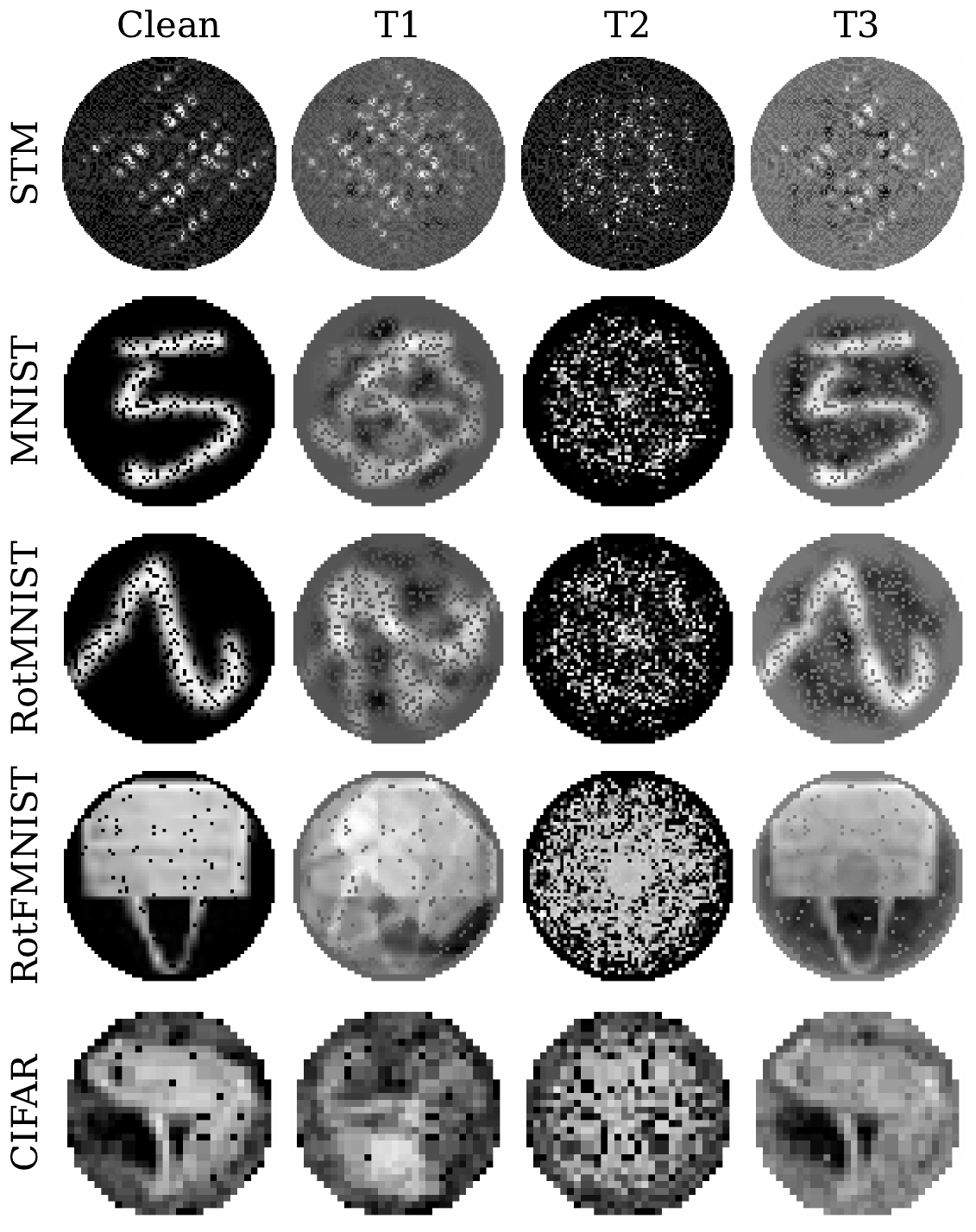}
    \caption{
    Representative examples from the datasets considered in this work (STM, MNIST, RotMNIST, RotFMNIST, and CIFAR). The first column shows a clean image, and the remaining columns show the corresponding transformed versions under T1, T2, and T3, respectively, as defined in Sec.~\ref{sec:transformations_theory}. Additional dataset details are provided in Appendix~\ref{app:num_setup}.
    }
    \label{fig:transformed_figure}
\end{figure}

\subsubsection{Readout modification}
\label{sec:architectural}

Instead of modifying the training data, the model’s readout can be adapted by composing the readout with a projector that suppresses selected Fourier modes. Here, we consider the specific intervention of suppressing the $m=0$ Fourier mode, i.e., the component associated with ring-average intensity. In the Fourier basis, \(m=0\) corresponds to the computational basis state
\(
\ket{0_{n_{\mathrm{orb}}}}=\ket{0}^{\otimes n_{\mathrm{orb}}}.
\)
We denote the associated orbital projector by
\begin{equation}
\Pi^{\mathrm{orb}}_0 := \ket{0_{n_{\mathrm{orb}}}}\!\bra{0_{n_{\mathrm{orb}}}} .
\end{equation}
For each class readout observable \(M_j := Z_j \otimes I_{\mathrm{orb}}\), we replace the standard measurement by
\begin{equation}
\tilde{M}_j \;:=\; Z_j \otimes \big(I_{\mathrm{orb}}-\Pi^{\mathrm{orb}}_0\big),
\end{equation}
where \(\big(I_{\mathrm{orb}}-\Pi^{\mathrm{orb}}_0\big)\ket{0_{n_{\mathrm{orb}}}}=0\).

\begin{table*}[t]
\setlength{\tabcolsep}{6pt}
\renewcommand{\arraystretch}{1.15}
\small

\noindent{(a) Train on clean, test on transformed data.}

\vspace{2pt}
{\centering
\begin{tabular}{lccccc}
\toprule
Test & STM & MNIST & RotMNIST & RotFMNIST & CIFAR \\
\midrule
Clean & $0.98 \pm 0.01$ & $0.84 \pm 0.01$ & $0.85 \pm 0.01$ & $0.72 \pm 0.01$ & $0.76 \pm 0.01$  \\

T1    & $0.98 \pm 0.01 (+0.00)$  & $0.84 \pm 0.01 (+0.00)$  & $0.85 \pm 0.01 (+0.00)$ & $0.72 \pm 0.01 (+0.00)$ & $0.76 \pm 0.01 (+0.00)$ \\
T2    & $0.84 \pm 0.02 (-0.14)$ & $0.35 \pm 0.01 (-0.49)$ & $0.37 \pm 0.02 (-0.48)$ & $0.42 \pm 0.03 (-0.30)$ & $0.69 \pm 0.01 (-0.07)$ \\
T3    & $0.82 \pm 0.02 (-0.17)$ & $0.55 \pm 0.01 (-0.29)$ & $0.50 \pm 0.01 (-0.35)$ & $0.43 \pm 0.02 (-0.29)$ & $0.72 \pm 0.02 (-0.04)$ \\
\bottomrule
\end{tabular}\par
}

\vspace{6pt}

\noindent{(b) Train on transformed data, test on clean.}

\vspace{2pt}
{\centering
\begin{tabular}{lccccc}
\toprule
Train & STM & MNIST & RotMNIST & RotFMNIST & CIFAR \\
\midrule
Clean & $0.98 \pm 0.01$ & $0.84 \pm 0.01$ & $0.85 \pm 0.02$ & $0.72 \pm 0.01$ & $0.76 \pm 0.01$  \\
T1    & $0.98 \pm 0.01 (+0.00)$ & $0.85 \pm 0.01 (+0.01)$ & $0.85 \pm 0.01 (+0.00)$ & $0.72 \pm 0.01 (+0.00)$  & $0.76 \pm 0.01 (+0.00)$  \\
T2    & $0.93 \pm 0.01 (-0.05)$ & $0.59 \pm 0.01 (-0.25)$ & $0.59 \pm 0.01 (-0.26)$ & $0.56 \pm 0.02 (-0.16)$ & $0.70 \pm 0.01 (-0.06)$  \\
T3    & $0.83 \pm 0.04 (-0.15)$ & $0.71 \pm 0.03 (-0.13)$ & $0.72 \pm 0.03 (-0.13)$ & $0.54 \pm 0.05 (-0.18)$  & $0.67 \pm 0.02 (-0.09)$ \\
\bottomrule
\end{tabular}\par
}

\caption{All reported values are mean test accuracies over five runs, evaluated using the best weights from each run, and are reported as mean $\pm$ one standard deviation. Values in parentheses report the performance change relative to the corresponding clean baseline for the same dataset, i.e., the difference in accuracy compared to training and evaluating on clean data. (a) The model is trained on clean data only and evaluated on the transformed test sets. (b) The model is trained on the transformed training sets and evaluated only on the clean test set.}
\label{tab:accessible_vs_exploited}
\end{table*}

\section{Numerical setup}
\label{sec:numerical_setup}

We briefly summarize the numerical setup used to evaluate the input transformations introduced in Sec.~\ref{sec:transformations_theory} and the robustness of the rotationally equivariant model under transfer attacks generated on classical surrogate models. Full details on datasets, training hyperparameters, and surrogate architectures are provided in Appendix~\ref{app:num_setup}.

\paragraph{Datasets.}
We evaluate our approach on five datasets: the STM dataset (10 classes) considered in Ref.~\cite{West_2024}, MNIST~\cite{lecun1998} and a rotated variant RotMNIST (each restricted to 9 classes), RotFMNIST (10 classes), a rotated variant of Fashion-MNIST~\cite{xiao2017FMNIST}, and a 2-class task derived from CIFAR~\cite{Krizhevsky2009CIFAR}. For MNIST and RotMNIST, we exclude the digit “9” to preserve rotational invariance at the label level.
Each dataset contains 7,500 training examples and 500 test examples.

\paragraph{Training-set variants.}
For each dataset, we consider the original training set (\emph{clean}) together with transformed variants obtained by applying each transformation from Sec.~\ref{sec:transformations_theory} to all training inputs while leaving the labels unchanged; these variants are denoted T1, T2, and T3. In addition, we include an adversarially trained variant obtained by augmenting the clean training set with adversarial examples generated by a linear surrogate (see Appendix~\ref{app:num_setup} for details).

\paragraph{Training.}
For each training-set variant, the rotationally equivariant model is trained for 10 epochs using a depth-100 circuit. All reported results are averaged over five independent training runs.

\paragraph{Adversarial examples.}
To evaluate transfer robustness, we generate adversarial examples using FGSM and PGD with four classical surrogate models: a linear classifier (LC), a multilayer perceptron (MLP), a convolutional neural network (CNN), and ResNet18~\cite{He2016}. For the LC and MLP surrogates, we use flattened inputs arranged in the same radial--orbital ordering as for the rotationally equivariant model. This choice enables a more direct comparison at the feature level. For CNN and ResNet18 we use the raw image inputs, since these architectures are designed to operate on spatially structured image data.
All surrogate models are trained exclusively on the clean training set. For each attack--surrogate combination, we generate 500 adversarial test images over a range of perturbation strengths $\varepsilon \in [0,0.3]$, measured in the $\ell_\infty$ norm. For CNN and ResNet18, adversarial examples are first generated in image space and then encoded before being passed to the target model.

\section{Results}
\label{sec:results}

In this section, we first analyze the effect of the input transformations introduced in Sec.~\ref{sec:transformations_theory} to infer which invariant features are used for classification. We then study the adversarial transfer robustness of the rotationally equivariant model and relate the observed robustness behavior to these feature-level findings. Finally, we examine adversarial training and show that robustness can be improved substantially by suppressing the $m=0$ Fourier mode.

\begin{figure*}[t]
    \centering
    \includegraphics[width=\linewidth]{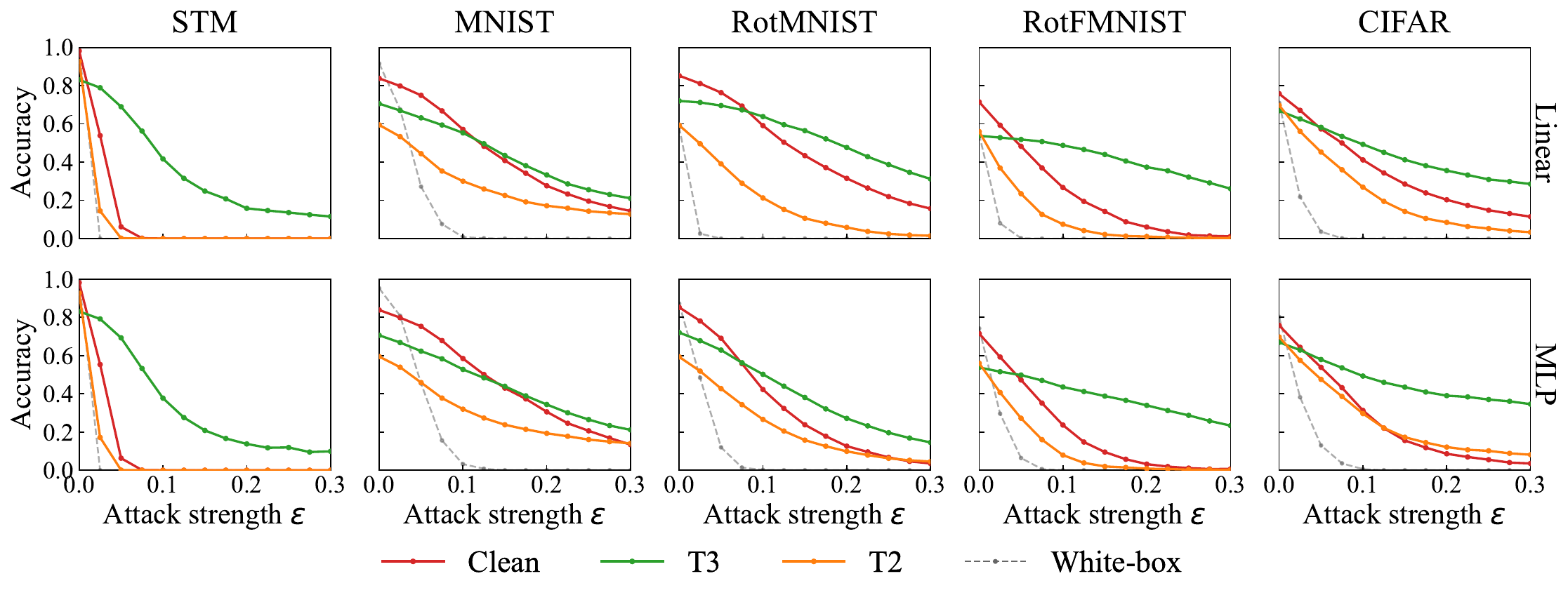}
    \caption{
    Test accuracy of the rotationally equivariant quantum model under transfer attacks across the STM, MNIST, RotMNIST, RotFMNIST, and CIFAR datasets for different training-data transformations. Rows 1--2 correspond to transfer attacks generated by clean-trained classical surrogate models, LC and MLP, respectively. Adversarial examples are generated using PGD on 500 test examples with $\ell_{\infty}$ perturbation strength $\varepsilon$. The gray dashed curve shows the corresponding surrogate white-box baseline, i.e., the accuracy of the surrogate model on its own adversarial examples. The green and orange curves show the mean test accuracy across five runs for target models trained on T3- and T2-transformed data, respectively. The red curve denotes the clean-trained baseline.
    }
    \label{fig:adv_robust}
\end{figure*}

\subsection{Accessible and utilized invariant features}
\label{sec:accessible_vs_used}

Table~\ref{tab:accessible_vs_exploited} summarizes two complementary train--test protocols across clean data and the transformations T1--T3. In panel (a), the model is trained on clean data and evaluated on transformed test sets. If performance remains close to the clean-test baseline, this suggests that the information preserved by the transformation constitutes a substantial part of the decision-relevant signal. In panel (b), the model is trained on transformed data and evaluated on the clean test set. If accuracy remains close to the clean-trained baseline, this suggests that the information retained under the transformation is already sufficient to learn a decision rule for the clean task. 

Training and evaluation on Clean versus T1 yields on average identical test accuracies across all datasets. This provides a direct numerical sanity check of the construction of T1 (see Appendix~\ref{app:transformations} for the details): although it visually scrambles the input, it acts only on components in the non-invariant subspace and therefore preserves the task-relevant information accessible to the rotationally equivariant model.

STM, RotFMNIST, and CIFAR exhibit a strong reliance on the information retained in T2. Although T2 removes much of the structure that appears visually salient to a human observer (see Figure~\ref{fig:transformed_figure}), clean-trained models retain comparatively high accuracy on T2. Moreover, models trained on T2 generalize to clean test data slightly better than T3-trained models, and substantially better in the case of STM. Together, these results suggest that ring-mean information accounts for a large fraction of the discriminative signal used by the model on these datasets.

For MNIST and RotMNIST, T3 retains more useful information for classification than T2, both when evaluating clean-trained models on transformed test data and when training directly on the transformed datasets. This matches the visual intuition that T3 preserves more of the salient digit structure that would appear class-relevant to a human observer. At the same time, the relatively weak performance even on T3 indicates that these features alone do not fully account for the discriminative signal present in the clean data.

Overall, while the model is restricted to rotation-invariant information, the particular invariant statistics it relies on vary substantially across datasets.

\begin{figure*}[t]
    \centering
    \includegraphics[width=\linewidth]{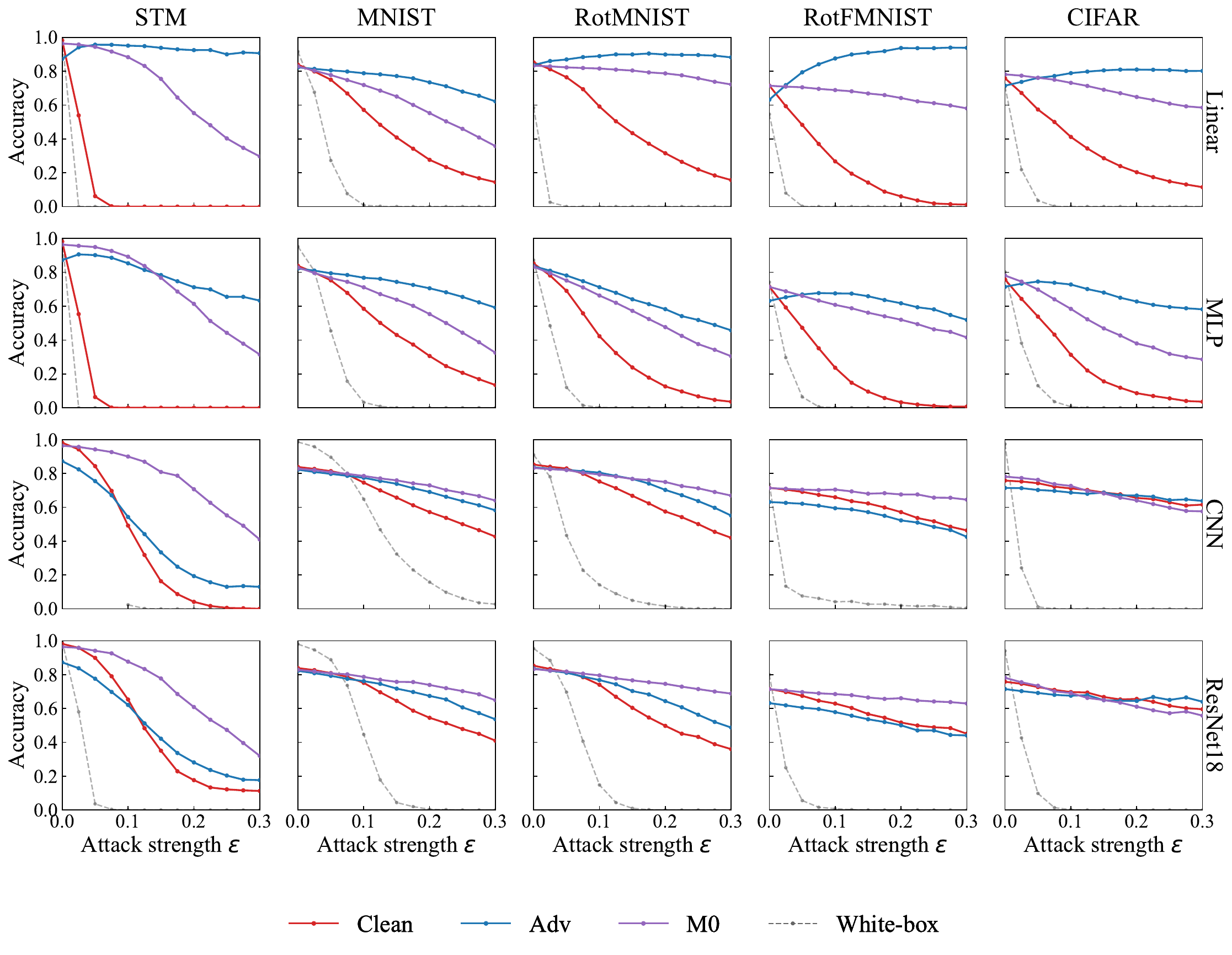}
   \caption{
   Test accuracy of the rotationally equivariant quantum model under adversarial training (Adv) and the architectural intervention (M0) across the STM, MNIST, RotMNIST, RotFMNIST, and CIFAR datasets.
    Rows 1--4 show transfer attacks generated by clean-trained classical surrogate models: LC, MLP, CNN, and ResNet18, respectively. Adversarial examples are generated using PGD on 500 test examples with $\ell_{\infty}$ perturbation strength $\varepsilon$. The gray dashed curve indicates the corresponding surrogate white-box baseline, i.e., the accuracy of the surrogate model on its own adversarial examples.
    The blue curve denotes the mean test accuracy of the rotationally equivariant model trained using adversarial training. The purple curve denotes the mean test accuracy of the same target model trained on clean data, with the $m=0$ mode projected out only at measurement. The red curve corresponds to the clean-trained baseline.}
    \label{fig:adv_training}
\end{figure*}

\subsection{Brittle and robust invariant features under transfer attacks}\label{sub:rob_features}

Since different datasets rely to different degrees on different parts of the accessible invariant feature space, we now ask whether these invariant features are equally robust under transfer attacks.
We focus on T2 and T3, as T1 was already shown to be equivalent to training on clean data and is therefore omitted from this analysis. To enable a direct feature-level comparison, we restrict this section to transfer attacks generated by the LC and MLP classical surrogates, which operate directly on inputs arranged in the same radial--orbital ordering as the target model. In this way, we reduce additional robustness effects that could arise from encoding or sampling mismatches.

Figure~\ref{fig:adv_robust} summarizes the robustness evaluation of the rotationally equivariant quantum model. The top row reports the accuracy on 500 test examples under PGD transfer attacks generated by the LC surrogate, while the bottom row shows the corresponding results for the MLP surrogate. We compare the clean-trained target model (red) with target models trained exclusively on T2 (orange) and T3 (green).

The model becomes substantially more robust when trained on T3 than when trained on clean data, although this comes at the cost of lower clean-test accuracy. By contrast, models trained on T2 are consistently less robust. Moreover, for STM, RotFMNIST, and CIFAR, the clean-trained model already exhibits poor robustness, at a level similar to T2. Notably, these are precisely the datasets that were previously identified as relying strongly on ring-average features for classification.

At first sight, this strong transferability may seem surprising, given that the surrogate can be as simple as a linear classifier and does not achieve high test accuracy on all datasets. However, the quantum models considered here can be viewed as linear models in feature-map Hilbert space~\cite{schuld2021,Schuld_2019}. Although the attacks are not applied directly in that space, they are performed on the encoded radial--orbital input representation, which is already closely aligned with the structure used by the target model. This likely contributes to why even LC-based surrogate attacks remain highly effective.

For the LC, this transfer mechanism can also be interpreted directly at the feature level. Appendix~\ref{app:lin_surro} shows that the learned LC is partially aligned with rotational invariance by primarily relying on ring-average information, and that this alignment is stronger for datasets with rotational augmentation. This supports the interpretation that the observed transferability arises, at least in part, because both models rely on similar ring-averaged statistics. More broadly, this supports the view that architectural differences alone do not prevent different models from relying on similar predictive features when they are trained on the same task and exploit the same informative structure in the data~\cite{goodfellow2015,ilyas2019}.

Overall, our results show that even within the restricted set of invariant features, the model can still rely on brittle statistics, in particular the ring-average features identified here. Equivariance alone therefore does not guarantee robustness in our setting.

\subsection{Improving robustness through targeted feature suppression}\label{sub:adv_rob}

The preceding results suggest two ways to improve the robustness of the rotationally equivariant model. The first is adversarial training, a standard approach in the robustness literature~\cite{madry2019,zhao2024,bai2021}. In our setting, the LC surrogate already yields strong attacks and therefore provides a computationally cheap source of adversarial examples. The second is an architectural intervention: training on clean data while projecting out the $m=0$ channel, corresponding to the ring-averaged intensities, which Section~\ref{sec:methods} identified as a likely source of non-robustness. This second strategy is particularly motivated by the fact that training on T3 also improves robustness, but at a substantial cost in clean-test accuracy; see Table~\ref{tab:accessible_vs_exploited} (b).

In Figure~\ref{fig:adv_training}, we compare adversarial training (blue), the $m=0$ projection strategy (purple), and the baseline model (red) under transfer attacks generated by four surrogate models: LC (top row), MLP (second row), CNN (third row), and ResNet18 (bottom row). Across all datasets, CNN and ResNet18 produce weaker attacks than LC and MLP, likely because they perturb the raw input space rather than the radial-orbital representation used by the target model.

Under LC-generated transfer attacks, the adversarially trained model attains higher accuracy on perturbed test inputs than on clean test inputs. Because 50\% of the training examples were adversarially perturbed, the learned decision boundary may be shifted toward adversarial rather than clean inputs. This is consistent with the well-known robustness--accuracy trade-off in adversarial training~\cite{tsipras2019}. In some cases, however, this trade-off can be mitigated by improved objectives or alternative training schemes~\cite{raghunathan2020,pang2022}.

This setting is also somewhat special, because the surrogate used for adversarial training is also the one used to evaluate robustness. For MLP, we observe a weaker but qualitatively similar effect. This suggests that adversarial examples generated by the LC are effective when the surrogate attacks directly in the radial-orbital representation used by the target model. The strongest improvement is observed for the datasets that were previously most vulnerable to the ring-average feature. Since the LC is partially aligned with ring-average information, this suggests that adversarial training in this setting primarily improves robustness against perturbations that act on that feature.

For the CNN and ResNet18 surrogates, adversarial training with examples generated by the LC yields a much less pronounced improvement compared to LC and MLP, especially when taking the associated trade-off with clean accuracy into account.

Projecting out the $m=0$ mode yields a consistent robustness improvement across all surrogates, while preserving clean-test accuracy better than adversarial training. For LC- and MLP-based attacks, the gain is smaller than that achieved by adversarial training. For CNN- and ResNet18-based attacks, however, it yields a larger robustness improvement than adversarial training on the LC-generated examples. Overall, this makes the 
$m=0$ projection a particularly effective method in our setting. Notably, it does not rely on data augmentation, but instead improves robustness by suppressing the brittle feature channel identified in our analysis.

\section{Discussion and Outlook}\label{sec:discussion}
Incorporating task symmetries into QML architectures has become a key route to obtaining trainable models and guiding circuit design. Here, we show that the introduced structure also enables a feature-level understanding of which input information the model retains and how this shapes downstream behavior such as transfer-based adversarial robustness.
Specifically, we characterize the rotationally equivariant quantum model of Ref.~\cite{West_2024} in terms of the invariant input statistics accessible to it, validate this picture through targeted input transformations, and use it to interpret the model’s transfer robustness under adversarial perturbations. We now discuss the main implications, limitations, and broader directions suggested by these findings.

A central question of this work was whether rotational equivariance biases the model toward features that are intrinsically more robust to adversarial perturbations. Our results suggest that this is not the case in general. Although equivariance renders the model insensitive to a large non-invariant subspace of perturbations, this blind subspace is geometrically structured and does not generally align with the perturbation directions exploited by adversarial attacks. As a result,  brittle and robust features can still coexist within the restricted class of rotation-invariant statistics. In particular, we find that simple invariant features, such as the ring-averaged intensities of the rotationally equivariant model, can remain highly predictive while being strongly non-robust, whereas richer invariant structures, such as higher-order angular correlation statistics, can be substantially more robust. At the same time, the equivariant structure—especially through its irreducible decomposition—provides a principled and targeted handle for controlling specific invariant features. In particular, we showed that projecting out the $m=0$ sector, which carries the ring-averaged intensities, increases the robustness of the model without any substantial loss in accuracy.

More broadly, this idea could be explored further by investigating in greater detail the signal contribution of the remaining symmetry channels, and by studying architectural strategies that either preserve them as separate feature streams or recombine them only at a later stage. Such designs could also support perturbation detection: if an attack selectively corrupts one channel, it may create inconsistencies across channels, so that cross-channel disagreement itself becomes a natural warning signal.

Further, our evaluation is restricted to a transfer-only adversarial setting.
This setting probes whether quantum and classical models rely on different feature representations under the same data distribution. For the datasets considered here, we find that even a simple linear surrogate can generate highly effective transfer perturbations when trained on a representation closely aligned with the model’s feature space. Thus, substantial transferability can arise without a highly expressive attacker. 
More broadly, these results suggest that obtaining genuinely different robustness behavior may require feature maps that make accessible features not readily available classically, whether because parts of the architecture are not classically simulable or because the data itself is inherently quantum.

For completeness, we also note that in the white-box setting—i.e., when the adversary can directly use the target model to craft adversarial examples—recent results for classical equivariant architectures report improved robustness, attributed to mechanisms such as gradient smoothing and orbit averaging~\cite{wang2025}. Although we leave the white-box setting to future work, related mechanisms may also be relevant for equivariant quantum models.

A further by-product of this feature-level analysis is that transformations acting within the insensitive subspace may also be useful for practical data obfuscation. We briefly discuss this perspective in Appendix~\ref{app:privacy}.

Lastly, while the feature analysis in this work is tied to the symmetry and input encoding of the rotationally equivariant model of Ref.~\cite{West_2024}, the same twirling-based reasoning extends to other strictly equivariant quantum models; what becomes case-specific is the interpretation of the resulting symmetry channels in terms of input-space features for a given encoding. Thus, this framework could be similarly used for a mechanistic analysis of other equivariant architectures.

Overall, our study shows how equivariance can enable a mechanistic, feature-level understanding of quantum models and thereby help explain properties such as adversarial robustness. Identifying the architectural and data-dependent conditions under which symmetry-based inductive biases lead to genuinely distinct robustness behavior remains an important direction for future work.

\section*{Acknowledgements}
We thank Joshua Combes and Maxwell West for valuable discussions and feedback.
MK acknowledges support from the Melbourne Research Scholarship.
The authors acknowledge the computational resources provided by the University of Melbourne’s Research Computing Service.

\bibliography{main}

\newpage
\onecolumngrid

\appendix
\section{Rotationally equivariant quantum model}
\label{app:rot_equiv_model_architecture}

For completeness, Figure~\ref{fig:rot_equiv_scheme} provides a schematic overview of the rotationally equivariant quantum model architecture introduced in Ref.~\cite{West_2024}.

\begin{figure}[h]
    \centering
    \includegraphics[width=0.65\linewidth]{rot_equiv_model.png}
    \caption{Architecture of the rotationally equivariant quantum model, adapted from Ref.~\cite{West_2024}. The input data are encoded into a quantum state via the encoding map $\mathcal{E}$ using a radial--orbital ordering. A quantum Fourier transform is then applied to the orbital register. Subsequently, only gates that are diagonal in the Fourier basis, here CZ gates, act on the orbital register, while the radial-register circuit remains fully expressive through arbitrary single-qubit rotations. Here, each $R(\theta_{i,j})$ block consists of $R_x(\theta_{i,j,0})$, $R_y(\theta_{i,j,1})$ and $R_z(\theta_{i,j,2})$
    rotations followed by nearest-neighbour CZ entangling gates. Finally, measurements are performed on the radial register.}
    \label{fig:rot_equiv_scheme}
\end{figure}

\section{Accessible input information via group twirling}
\label{group_twirl_to_input}

Let $\mathcal{G}$ be a finite group with unitary representation $R(g)$ acting on the input Hilbert space.
Assume (i) the variational unitary $U_{\theta}$ is $\mathcal{G}$-equivariant 
\begin{equation}
U_\theta R(g)=R(g)U_\theta \qquad \forall g\in \mathcal{G},
\end{equation}
and (ii) the readout observable $M$ is $\mathcal{G}$-invariant, i.e.
\begin{equation}
R(g)^\dagger M R(g)=M \qquad \forall g\in \mathcal{G} .
\end{equation}
Define the twirling channel
\begin{equation}
\mathcal T_\mathcal{G}(\rho):=\frac{1}{|\mathcal{G}|}\sum_{g\in \mathcal{G}} R(g)\rho R(g)^\dagger .
\end{equation}
Then, for any input state $\rho$,
\begin{align}
\Tr\!\left(M\,U_\theta\rho U_\theta^\dagger\right)
&=\frac{1}{|\mathcal{G}|}\sum_{g\in \mathcal{G}}\Tr\!\left(M\,U_\theta R(g)\rho R(g)^\dagger U_\theta^\dagger\right)\\
&=\Tr\!\left(M\,U_\theta \mathcal T_\mathcal{G}(\rho) U_\theta^\dagger\right),
\end{align}
where we used the assumptions of (i), (ii) and the linearity and cyclicity of the trace.
Hence the model cannot distinguish $\rho$ from its twirled counterpart $\mathcal T_\mathcal{G}(\rho)$.

We now apply this to the rotationally equivariant model of Ref.~\cite{West_2024}, where
$\mathcal{G}=\mathbb Z_{N_\phi}$ acts as cyclic shifts on the orbital register and trivially on the radial register.
Using the encoding of Ref.~\cite{West_2024}, the input image $x$ is represented by
\begin{equation}
\ket{\psi( x)}=\frac{1}{\|x\|}\sum_{r=0}^{N_r-1}\sum_{\phi=0}^{N_\phi-1} x_{r,\phi}\ket{r,\phi}\label{eq:fourier_state},
\end{equation}
We then transform to the Fourier basis on the orbital register, 
\begin{equation}
\ket{m}:=\frac{1}{\sqrt{N_\phi}}\sum_{\phi=0}^{N_\phi-1}\omega^{-m\phi}\ket{\phi},
\qquad \omega:=e^{2\pi i/N_\phi},
\end{equation}
so that
\begin{equation}
\ket{\psi_F( x)}=\bigl(\mathbb{I}_{\mathrm{rad}}\otimes \mathrm{QFT}^\dagger_{\mathrm{orb}}\bigr)\ket{\psi(x)}=\frac{1}{\|x\|}\sum_{r,m}\tilde x_{r,m}\ket{r,m},
\qquad
\tilde x_{r,m}=\frac{1}{\sqrt{N_\phi}}\sum_{\phi=0}^{N_\phi-1} x_{r,\phi}\,\omega^{-m\phi}.
\end{equation}

In this basis the group action $R(g):=I_{\mathrm{rad}}\!\otimes R_{\text{orb}}(g)$ is diagonal, with
\begin{equation}
R_{\text{orb}}(g)=\sum_{m=0}^{N_\phi-1}\omega^{mg}\ket m\!\bra m,
\end{equation}
and twirling the density matrix $\rho=\ket{\psi_F(x)}\!\bra{\psi_F(x)}$ yields
\begin{align}
\mathcal T_{\mathbb Z_{N_\phi}}(\rho)
&=\frac{1}{N_\phi}\sum_{g=0}^{N_\phi-1}(I_{\mathrm{rad}}\!\otimes R_{\text{orb}}(g))\,\rho\,(I_{\mathrm{rad}}\!\otimes R_{\text{orb}}(g))^\dagger\\
&=\frac{1}{\| x\|^2}\sum_{r,r'}\sum_{m}
\tilde x_{r,m}\tilde x^{*}_{r',m}\,\ket{r,m}\!\bra{r',m},
\label{eq:twirl_outcome}
\end{align}
where we used
\begin{align}
    \frac{1}{N_{\phi}}\sum_g   \omega^{(m'-m)g}= \delta_{m',m}.
\end{align}.

To interpret the twirled density matrix $\mathcal{T}_{\mathbb Z_{N_\phi}}(\rho)$ in the pixel basis $\ket{r,\phi}$, we move back from the Fourier basis to the orbital basis, i.e.,
\begin{align}
(\mathbb I_{\rm rad}\!\otimes\!\mathrm{QFT}_{\rm orb})\,\mathcal T_{\mathbb Z_{N_\phi}}(\rho)\,
(\mathbb I_{\rm rad}\!\otimes\!\mathrm{QFT}_{\rm orb}^\dagger)
&=\frac{1}{\| x\|^2}\sum_{r,r'}\sum_{m}
\tilde x_{r,m}\tilde x^{*}_{r',m}\,\frac{1}{N_{\phi}}\sum_{\phi,\phi'}\omega^{m(\phi-\phi')}\ket{r,\phi}\!\bra{r',\phi'}\notag\\
&=\frac{1}{\| x\|^2}\sum_{r,r'}\sum_{m}\frac{1}{N_{\phi}}\sum_{\alpha,\beta}
 x_{r,\alpha}\, x^{*}_{r',\beta}\,\omega^{-m(\alpha-\beta)}\,\frac{1}{N_{\phi}}\sum_{\phi,\phi'}\omega^{m(\phi-\phi')}\ket{r,\phi}\!\bra{r',\phi'}\notag\\
&=\frac{1}{N_{\phi}^2\| x\|^2}\sum_{r,r'}\sum_{m}\sum_{\alpha,\beta}\sum_{\phi,\phi'}
 x_{r,\alpha}\, x^{*}_{r',\beta}\,\omega^{-m\big((\alpha-\beta)-(\phi-\phi')\big)}\,\ket{r,\phi}\!\bra{r',\phi'}\notag\\
& \overset{1)}{=}\frac{1}{N_{\phi}^2\| x\|^2}\sum_{r,r'}\sum_{m}\sum_{\alpha,\beta}\sum_{\phi,\Delta\phi}
 x_{r,\alpha}\, x^{*}_{r',\beta}\,\omega^{-m\big(\alpha-\beta-\Delta\phi\big)}\,\ket{r,\phi}\!\bra{r',\phi-\Delta\phi}\notag\\
&=\frac{1}{N_{\phi}\| x\|^2}\sum_{r,r'}\sum_{\alpha,\beta}\sum_{\phi,\Delta\phi}
 x_{r,\alpha}\, x^{*}_{r',\beta}\,
\delta_{\alpha-\beta-\Delta\phi\equiv 0}\,
\ket{r,\phi}\!\bra{r',\phi-\Delta\phi}\notag\\
&=\frac{1}{N_{\phi}\| x\|^2}\sum_{r,r'}\sum_{\alpha}\sum_{\phi,\Delta\phi}
 x_{r,\alpha}\, x^{*}_{r',\alpha-\Delta\phi}\,
\ket{r,\phi}\!\bra{r',\phi-\Delta\phi},\label{eq:final_outcome}
\end{align}
where in 1) we used $\Delta\phi=(\phi-\phi')\bmod N_{\phi}$, and all orbital indices are understood modulo $N_{\phi}$.

Equation~\eqref{eq:final_outcome} makes explicit which input information remains accessible under $\mathbb Z_{N_\phi}$-equivariance together with an invariant readout: the model cannot depend on the \emph{absolute} angular position, but only on information encoded in rotation-invariant correlations, i.e., quantities that depend on relative angular offsets $\Delta\phi$ rather than absolute angles. Indeed, after transforming back to the pixel (orbital) basis, the twirled state is governed by coefficients of the form
\begin{align}
C_{r,r'}(\Delta\phi)\;:=\;\sum_{\alpha=0}^{N_\phi-1} x_{r,\alpha}\,x^{*}_{r',\alpha-\Delta\phi},
\qquad \Delta\phi\in\mathbb Z_{N_\phi},\label{eq:corr}
\end{align}
i.e., circular correlations between rings at radii $r$ and $r'$ as a function of the relative angular offset $\Delta\phi$ (indices understood modulo $N_\phi$). This dependence on $\Delta\phi$ alone is precisely what enforces invariance: a global rotation $x_{r,\phi}\mapsto x_{r,\,(\phi+g)\bmod N_{\phi}},\quad g\in\mathbb Z_{N_\phi}$ leaves $C_{r,r'}(\Delta\phi)$ unchanged.

\section{Input transformations}
\label{app:transformations}

We introduce three input transformations to numerically support and probe the twirling-based feature analysis of the previous section. Throughout, we use the following notation. Let $\mathcal{D}=\{(x_i,y_i)\}_{i=1}^N$ and represent each image $x$ by its ring samples
\[
x_r := (x_{r,0},\dots,x_{r,N_\phi-1})^\top \in \mathbb{R}^{N_\phi},
\qquad r\in\{0,\dots,N_r-1\},
\]
where $N_\phi=2^{n_{\mathrm{orb}}}$ and $N_r=2^{n_{\mathrm{rad}}}$.

\subsubsection*{Transformation 1 (orthogonal circulant scrambling).}
For each image $x$, we construct a transformed image $x^{(1)}$ by applying the \emph{same} random orthogonal circulant transform along the angular direction on every ring vector $x_r$:
\begin{enumerate}
\item Sample a random diagonal phase vector $\lambda\in\mathbb{C}^{N_\phi}$ with $|\lambda_m|=1$, conjugate symmetry $\lambda_{N_\phi-m}=\lambda_m^*$, $\lambda_{N_{\phi}/2}\in\{\pm 1\}$ and $\lambda_0=1$, so that the resulting transform is real-valued.
\item Define the corresponding orthogonal circulant matrix
\begin{equation}
O(\lambda) := F^\dagger \,\mathrm{diag}(\lambda)\, F \in \mathbb{R}^{N_\phi\times N_\phi},
\end{equation}
where $F$ denotes the unitary discrete Fourier transform on $\mathbb{C}^{N_\phi}$.
\item Apply the same $O(\lambda)$ to all rings,
\begin{equation}
x_r^{(1)} := O(\lambda)\,x_r \qquad \forall\, r\in\{0,\dots,N_r-1\}.
\end{equation}
\end{enumerate}
The transformed image is then given by $x^{(1)} := \{x_r^{(1)}\}_{r=0}^{N_r-1}$.
By construction, this transformation preserves the rotation-invariant correlation features
$C_{r,r'}(\Delta\phi)$.
Let $S_{\Delta\phi}$ denote the cyclic shift on $\mathbb{R}^{N_\phi}$ defined by
$(S_{\Delta\phi}x_r)_\alpha := x_{r,\alpha-\Delta\phi\;(\mathrm{mod}\,N_\phi)}$.
Then
\[
C_{r,r'}(\Delta\phi)=\sum_{\alpha=0}^{N_\phi-1} x_{r,\alpha}\,x_{r',\alpha-\Delta\phi}
= x_r^\top S_{\Delta\phi} x_{r'}.
\]

Under the transformation $x_r^{(1)} = O(\lambda) x_r$ (with the \emph{same} circulant $O(\lambda)$ for all $r$), we obtain
\[
C'_{r,r'}(\Delta\phi)
= (x_r^{(1)})^\top S_{\Delta\phi} x_{r'}^{(1)}
= x_r^\top O(\lambda)^\top S_{\Delta\phi} O(\lambda)\, x_{r'} .
\]
Since $O(\lambda)$ is circulant, it commutes with every cyclic shift: $O(\lambda)S_{\Delta\phi}=S_{\Delta\phi}O(\lambda)$.
Moreover, $O(\lambda)^\top$ is also circulant and therefore commutes with $S_{\Delta\phi}$ as well.
Hence
\[
O(\lambda)^\top S_{\Delta\phi} O(\lambda) = S_{\Delta\phi} O(\lambda)^\top O(\lambda) = S_{\Delta\phi},
\]
where we used orthogonality $O(\lambda)^\top O(\lambda)=I$. Consequently, $C'_{r,r'}(\Delta\phi)=C_{r,r'}(\Delta\phi)$ for all
$r,r'$ and $\Delta\phi$.

\subsubsection*{Transformation 2 (ring-wise permutation).}
For each image $x$, we construct for each ring $r$ a transformed ring vector $x_r^{(2)}$ as follows: Sample an independent permutation matrix $P_r\in\mathbb{R}^{N_\phi\times N_\phi}$ uniformly from the set of permutation matrices and permute the orbital samples,
\begin{equation}
x_r^{(2)} := P_r x_r .
\end{equation}
The transformed image is then given by \(x^{(2)} := \{x_r^{(2)}\}_{r=0}^{N_r-1}\).
By construction, \(x^{(2)}\) preserves the ring-wise means, while scrambling the angular ordering so that, in general, the full correlation features
\(C_{r,r'}(\Delta\phi)\) are not preserved.

\subsubsection*{Transformation 3 (ring-wise mean removal).}
For each image $x$, we construct a transformed image $x^{(3)}$ by removing the ring-mean on every ring $r$.
Define the ring-wise mean as
\[
\bar x_r := \frac{1}{N_\phi}\mathbf{1}^\top x_r,
\qquad r\in\{0,\dots,N_r-1\}.
\]
The transformation is then given by subtracting the ring mean $\bar x_r$ from each ring vector $x_r$:
\begin{equation}
x^{(3)}_r = x_r - \bar x_r\,\mathbf{1},
\qquad \forall\, r\in\{0,\dots,N_r-1\},
\end{equation}
and $x^{(3)} := \{x^{(3)}_r\}_{r=0}^{N_r-1}$.
By construction, each transformed ring has zero mean, $\mathbf{1}^\top x^{(3)}_r = 0$.
After mean removal $x^{(3)}_r=x_r-\bar x_r\mathbf 1$, the corresponding correlations
$C'_{r,r'}(\Delta\phi):=(x^{(3)}_r)^\top S_{\Delta\phi}x^{(3)}_{r'}$ satisfy
\begin{align}
C'_{r,r'}(\Delta\phi)
&=(x_r-\bar x_r\mathbf 1)^\top S_{\Delta\phi}(x_{r'}-\bar x_{r'}\mathbf 1)\notag\\
&=x_r^\top S_{\Delta\phi}x_{r'}
-\bar x_{r'}\,x_r^\top S_{\Delta\phi}\mathbf 1
-\bar x_r\,\mathbf 1^\top S_{\Delta\phi}x_{r'}
+\bar x_r\bar x_{r'}\,\mathbf 1^\top S_{\Delta\phi}\mathbf 1\notag\\
&=C_{r,r'}(\Delta\phi)
-\bar x_{r'}\,\mathbf 1^\top x_r
-\bar x_r\,\mathbf 1^\top x_{r'}
+\bar x_r\bar x_{r'}\,\mathbf 1^\top \mathbf 1\notag\\
&=C_{r,r'}(\Delta\phi)-N_\phi\,\bar x_r\,\bar x_{r'},
\label{eq:mean_removed_corr_simple}
\end{align}
where we used $C_{r,r'}(\Delta\phi)=x_r^\top S_{\Delta\phi}x_{r'}$, the invariance $S_{\Delta\phi}\mathbf 1=\mathbf 1$, and $\mathbf 1^\top x_r = N_\phi \bar x_r$.
Thus, at the level of the unnormalized correlations, Transformation~3 preserves the
\(\Delta\phi\)-dependence of \(C_{r,r'}(\Delta\phi)\) up to a constant offset term.
After amplitude encoding, the resulting state is additionally rescaled by the new norm
\(\|x^{(3)}\|\).

\section{Numerical supplement}\label{app:num_setup}

This appendix provides further implementation details on the datasets, model architectures, and adversarial attacks used in this work.

\subsubsection{Datasets, preprocessing, and input encoding.}
Table~\ref{tab:dataset_overview} summarizes the datasets used in this work—STM, MNIST, RotMNIST, RotFMNIST, and CIFAR—as well as their respective configurations. The STM dataset is constructed following the setup of Refs.~\cite{Usman_STM, West_STM, Usman_2017}. RotMNIST denotes a rotated variant of MNIST~\cite{lecun1998} constructed for this work, and RotFMNIST is defined analogously from the original Fashion-MNIST dataset~\cite{xiao2017FMNIST}. Finally, CIFAR refers to a binary classification task derived from the original CIFAR dataset~\cite{Krizhevsky2009CIFAR}.

For MNIST and RotMNIST we excluded the class "9" to ensure rotational invariance at the label level.
For all datasets, we first generated grayscale input images, with the resolution adapted to match the input encoding used later. For STM, the dataset was constructed by applying Gaussian noise and rotations in steps of $45^\circ$ to an original set containing one image per class, similar to the procedure described in Ref.~\cite{West_2024}. For RotMNIST and RotFMNIST, we likewise introduced rotations in steps of $45^\circ$, consistent with the 
$\mathbbm{Z}_8$-invariance of our model. From these images, we then generate the encoded input states (flattened encoded vectors) as described in Sec.~\ref{sec:rotequiv}. For all datasets, we fixed $n_{\text{orb}}=3$, corresponding to $\mathbbm{Z}_8$-invariance, and chose $n_{\text{rad}}$ such that the number of classes and the target image resolution were appropriately represented. 

\begin{table*}[t]
\centering
\renewcommand{\arraystretch}{1.6}
\setlength{\tabcolsep}{6pt}

\begin{tabular*}{\textwidth}{@{\extracolsep{\fill}}lccccc@{}}
\hline
\textbf{Property} & \textbf{STM} & \textbf{MNIST} & \textbf{RotMNIST} & \textbf{RotFMNIST} & \textbf{CIFAR} \\
\hline
Classes & 10 & 9 & 9 & 10 & 2 \\
Resolution & $128\times128$ & $64\times64$ & $64\times64$ & $64\times64$ & $32\times32$ \\
Input format & grayscale & grayscale & grayscale & grayscale & grayscale \\
Data augmentation 
& \parbox[c]{2.8cm}{\centering Gaussian noise + 8 discrete rotations}
& none
& \parbox[c]{2.1cm}{\centering 8 discrete rotations}
& \parbox[c]{2.1cm}{\centering 8 discrete rotations}
& none \\
Encoding $(n_{\mathrm{rad}}, n_{\mathrm{orb}})$ & $(10,3)$ & $(9,3)$ & $(9,3)$ & $(10,3)$ & $(8,3)$ \\
Train / test size & 7500 / 500 & 7500 / 500 & 7500 / 500 & 7500 / 500 & 7500 / 500 \\
Comments 
& --
& \parbox[c]{1.9cm}{\centering digit ``9'' excluded}
& \parbox[c]{1.9cm}{\centering digit ``9'' excluded}
& --
& \parbox[c]{2.4cm}{\centering binary task:\\ frog vs.\ airplane} \\
Examples (one per class)
& \parbox[t][2.25cm][t]{2.25cm}{\centering\vspace{0pt}\includegraphics[width=2.7cm,keepaspectratio]{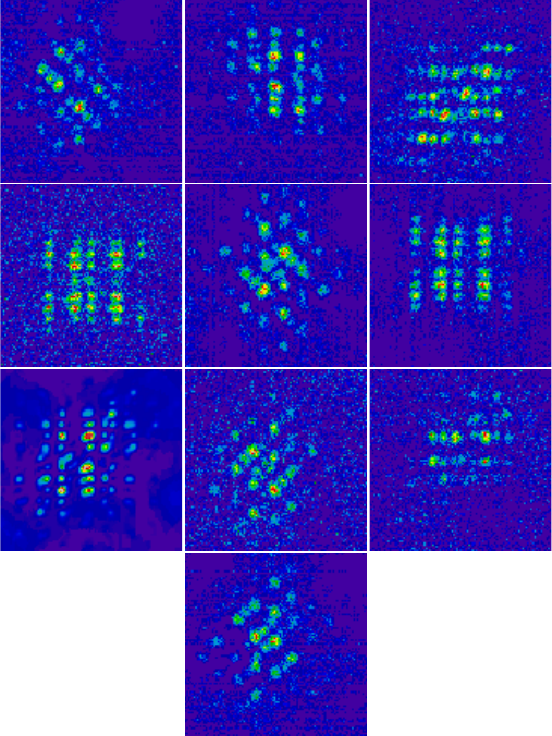}}
& \parbox[t][2.25cm][t]{2.25cm}{\centering\vspace{0pt}\includegraphics[width=2.7cm,keepaspectratio]{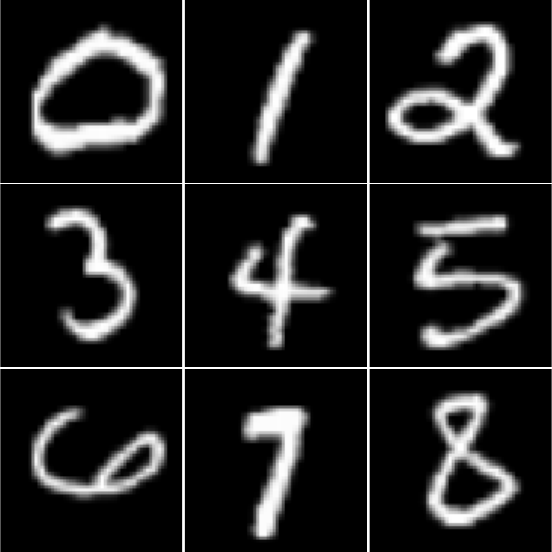}}
& \parbox[t][2.25cm][t]{2.25cm}{\centering\vspace{0pt}\includegraphics[width=2.7cm,keepaspectratio]{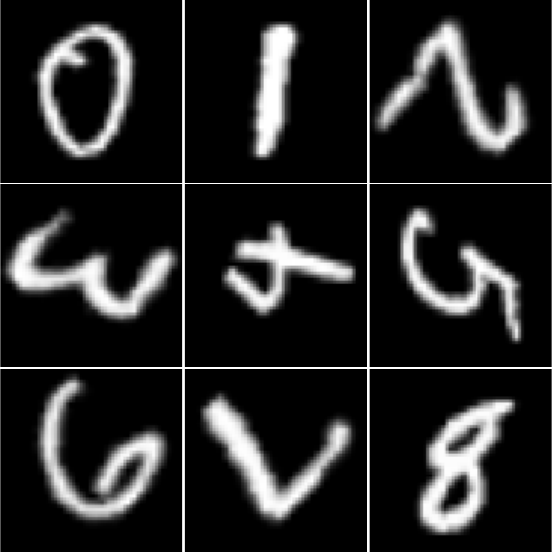}}
& \parbox[t][2.25cm][t]{2.25cm}{\centering\vspace{0pt}\includegraphics[width=2.7cm,keepaspectratio]{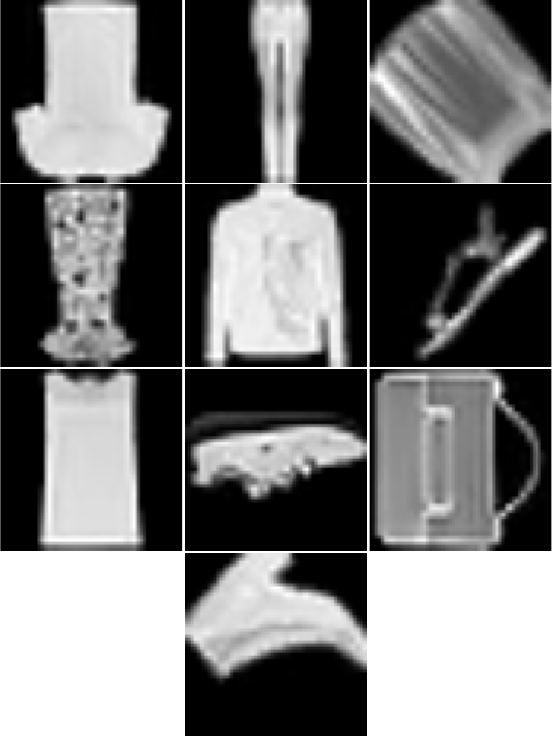}}
& \parbox[t][2.25cm][t]{2.25cm}{\centering\vspace{0pt}\includegraphics[width=2.7cm,keepaspectratio]{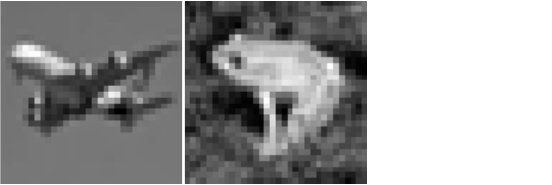}}\\[0.25cm]
\rule{0pt}{1.3cm}\\
\hline
\end{tabular*}
\caption{Overview of the datasets used in this work. Example images are shown in ascending class order, read from left to right and top to bottom, with one representative image per class. The STM images are displayed in color for better visualization, although they are also used in grayscale in the experiments. The listed image resolutions refer to the original grayscale images before encoding according to Eq.~\ref{enc_state}.}
\label{tab:dataset_overview}
\end{table*}

\subsubsection{Models and training.}

\paragraph{Rotationally equivariant quantum model.}
We adopt the rotationally equivariant quantum model introduced in Ref.~\cite{West_2024}. Across all datasets, we use a circuit depth of 100 layers, a batch size of 16, and a learning rate of $3\times 10^{-3}$. Each model is trained for 10 epochs, and all experiments are repeated over five random seeds.

\paragraph{Classical surrogate models.} We consider four classical surrogate models: a linear classifier (LC), a multilayer perceptron (MLP), a convolutional neural network (CNN), and a ResNet18 model. The LC and MLP operate on the radial-orbital ordered representation, whereas the CNN and ResNet18 are trained on the grayscale image inputs.

The LC consists of a single linear layer mapping the input dimension $D$ directly to the number of classes $C$. The MLP consists of one hidden layer of width 256 with ReLU activation, followed by the output layer. The CNN is a custom convolutional architecture with multiple convolutional layers, batch normalization, ReLU activations, intermediate max-pooling, and a fully connected classification head. The ResNet surrogate is based on a standard ResNet18 architecture, see Ref.~\cite{He2016}, with randomly initialized weights; the first convolutional layer was adapted to a single grayscale input channel, and the final fully connected layer was replaced to match the number of output classes.

All surrogate models were implemented in PyTorch and trained using the cross-entropy loss and the AdamW optimizer with batch size 128 and weight decay $10^{-4}$.

\subsubsection{Adversarial test examples.}
Using each classical surrogate model, we constructed two adversarially perturbed versions of the test set, one using FGSM and one using PGD as described in Sec.~\ref{sec:adv_attacks_theory}. For each surrogate, adversarial examples were generated using the checkpoint with the highest clean test accuracy. For each selected checkpoint, attacks were generated over the perturbation range $\varepsilon \in [0,0.3]$ in increments of $0.025$, with $\varepsilon$ measured in the $\ell_\infty$ norm. For PGD, we used $T=10$ projected gradient steps and a step size $\alpha=\max(\varepsilon/T,10^{-3})$. For the CNN and ResNet surrogates, attacks were generated in the raw input space and the resulting adversarial examples were then mapped into the encoded input representation used by the rotationally equivariant model.

\subsubsection{Adversarial training.}
For each dataset, we constructed a mixed adversarial training set by selecting 50\% of the original training samples uniformly at random, while leaving the remaining samples unchanged. For each selected sample, we replace it with an adversarial counterpart using a FGSM attack on the LC surrogate model, with an individual perturbation strength $\varepsilon$ drawn uniformly from the interval $[0.05, 0.2]$. The perturbations were projected onto the corresponding $\ell_\infty$ ball.

\section{Practical obfuscation of input data}
\label{app:privacy}

\begin{figure}
    \centering
    \includegraphics[width=0.5\linewidth]{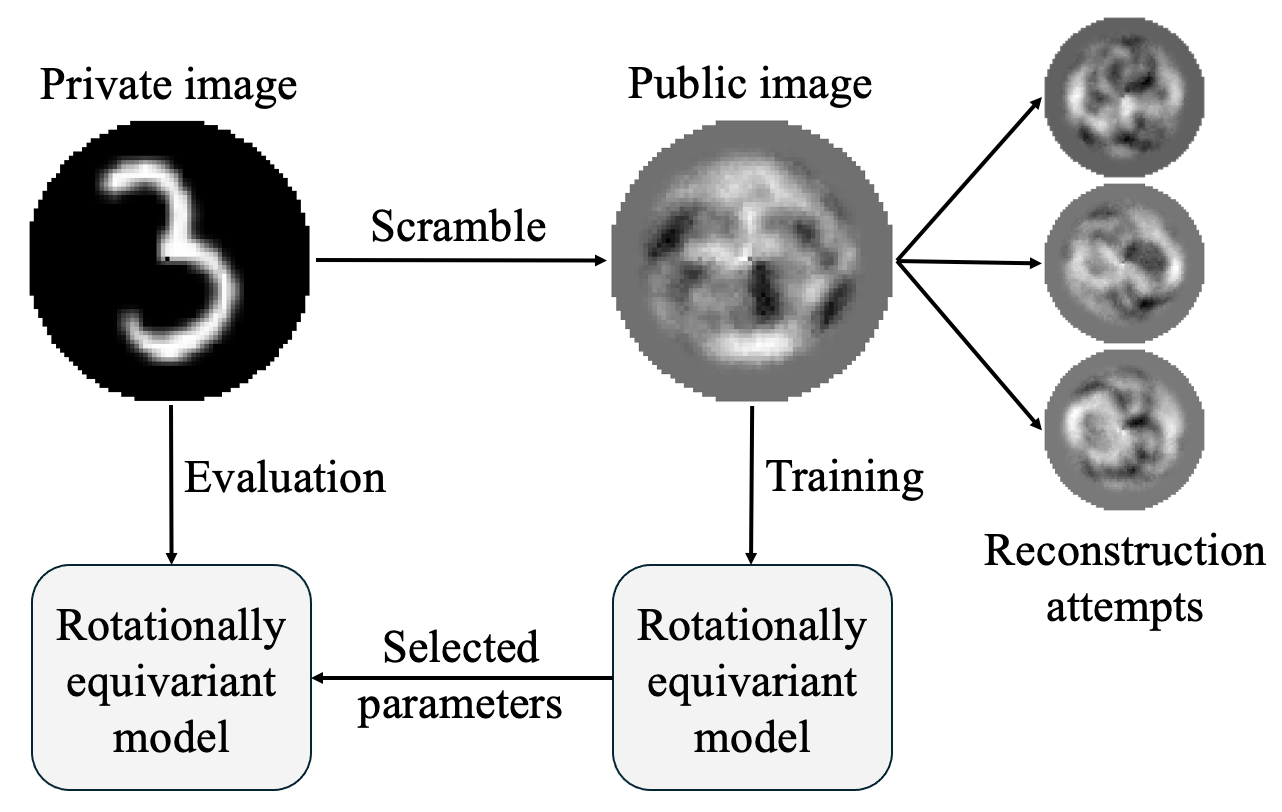}
    \caption{A private image $x$ is mapped to a public scrambled image $x_r' = O(\lambda_{\mathrm{key}})\,x_r$. In reconstruction attempts, different guessed keys $\tilde{\lambda}$ generally yield different candidate preimages that are all consistent with the same released image $x'$. Since the scrambling preserves the rotation-invariant statistics accessible to the rotationally equivariant model, training on scrambled data yields task utility comparable to that obtained from clean private images. After training, the selected model configuration, including learned parameters and hyperparameters, can be transferred to the private setting for evaluation on private images.}
    \label{fig:privacy_setup}
\end{figure}

Our feature analysis suggests a practical obfuscation mechanism: once sensitive and insensitive perturbation directions are identified, images can be deliberately perturbed in directions to which the model is largely insensitive, thereby degrading their visual interpretability while preserving much of the information used for prediction. This is related to prior work on transformation-based privacy and obfuscation via geometry-preserving mappings~\cite{chen_2008,chen_2011,liu_2006}. In particular, the results in Table~\ref{tab:accessible_vs_exploited} suggest that T1 (see Sec.~\ref{sec:transformations_theory}) can serve as a practical obfuscation scheme.

Consider the following threat model: (i) only scrambled images $x_r' = O(\lambda_{\mathrm{key}})\,x_r$ are released; (ii) the original images $x$ are never shared (hence no paired observations $(x,x')$ are available); and (iii) the per-image key $\lambda_{\mathrm{key}}$ is unknown to the adversary. Under this threat model, training on the public representations $x'$ can retain comparable utility to training on the clean inputs, because T1 preserves the rotation-invariant correlations accessible to the rotationally equivariant model (see Table~\ref{tab:accessible_vs_exploited}).

At the same time, recovering $x$ from a single released image $x'$ is underdetermined without additional side information. Any reconstruction attempt must effectively guess a key $\tilde{\lambda} \in \Lambda$ (from the set of admissible keys). For any $\tilde{\lambda} \in \Lambda$, the candidate preimage
\begin{equation}
x_r(\tilde{\lambda}) := O(\tilde{\lambda})^{\top} x_r'
\end{equation}
satisfies $O(\tilde{\lambda})\,x_r(\tilde{\lambda})=x_r'$. Hence many distinct candidates are consistent with the same released image $x'$, and therefore, the released image alone does not determine a unique reconstruction.

The degree of visual obfuscation, and the practical difficulty of reconstruction under this threat model, depends on the angular discretization $\mathbbm{Z}_{N_{\phi}}$ and can be expected to increase with $N_{\phi}$ (as this increases the dimension of the key $\lambda_{\mathrm{key}}$). We emphasize that this does not constitute a formal privacy guarantee (e.g., differential privacy~\cite{dwork_2006}); establishing such guarantees would require further analysis. Nevertheless, the scheme in Fig.~\ref{fig:privacy_setup} illustrates how data may be shared in scrambled form for external training, hyperparameter tuning, or benchmarking without directly exposing the original pixel-level content $x$.
Table~\ref{tab:accessible_vs_exploited} also shows that, for some datasets (e.g., STM), T2 can retain a substantial fraction of the task utility based on empirical observation. In such cases, if the collection $\{P_r\}_r$ is unknown, the random ring-wise permutations of T2 can make recovery of the original angular ordering substantially more ambiguous in practice than using T1.

\section{Ring-invariant structure in the linear surrogate model}
\label{app:lin_surro}

 Consider the feature vector $x= (x_{r,\phi})_{r,\phi}$ 
defined in Eq.~\eqref{enc_state}.
Given a weight matrix $W \in \mathbb{R}^{n_{\mathrm{classes}} \times (N_r N_\phi)}$ and a bias vector $b \in \mathbb{R}^{n_{\mathrm{classes}}}$, a linear multiclass classifier assigns a score to each class $c$ via
\begin{equation}
f_c(x)
= \sum_{r=0}^{N_r-1}\sum_{\phi=0}^{N_\phi-1} w_{c,r,\phi}\, x_{r,\phi} + b_c,
\qquad c=1,\dots,n_{\mathrm{classes}}.
\end{equation}
The predicted label is then given by
\begin{equation}
\hat y(x) = \argmax_{c \in \{1,\dots,n_{\mathrm{classes}}\}} f_c(x).
\end{equation}

\begin{table}[t]
\centering
\begin{tabular}{lccccc}
\hline
 & CIFAR & MNIST & RotFMNIST & RotMNIST & STM \\
\hline
$S_{\text{ring}}$ & 0.127 & 0.162 & 0.401 & 0.306 & 0.463 \\
\hline
\end{tabular}
\caption{Ring-invariance score $S_{\text{ring}}$ of the best linear classifier for each dataset.}
\label{tab:ring_score}
\end{table}

A sufficient condition for rotation-invariant predictions is that each class score is itself invariant under rotations. For the linear classifier, this is equivalent to
\begin{align}
f_c(x) = f_c(R(g)x) \ \forall\, x, g
\quad \Longleftrightarrow \quad
w_{c,r,\phi}=w_{c,r,\phi+g\,(\mathrm{mod}\,N_\phi)}
\ \forall\, c,r,\phi,g .
\end{align}
where $(R(g)x)_{r,\phi}=x_{r,\phi+g\,(\text{mod}\, N_\phi)}$ denotes the action of a discrete rotation by $g$ on the orbital index.
This is equivalent to the weights being constant along each ring, i.e., $w_{c,r,\phi}\equiv w_{c,r}$, which yields
\begin{equation}
  f_c(x)=\sum_{r} w_{c,r}\sum_{\phi} x_{r,\phi}
  = N_\phi \sum_r w_{c,r}\,\bar x_r,
\end{equation}
with $\bar x_r := \frac{1}{N_\phi}\sum_{\phi} x_{r,\phi}$, and for simplicity we set \(b_c = 0\).

Note that this is already a strong notion of invariance, as it requires the model to assign exactly the same class scores to an input and all of its rotations. For rotation-invariant predictions, however, it would already be sufficient that the logits are preserved up to a class-independent additive shift, since this leaves the $\arg\max$ unchanged.

To assess how closely the learned linear classifier aligns with a classifier that depends only on ring-averaged features, we take the best-performing weight matrix $W$ for each dataset and project it onto the subspace of weights that are constant along the orbital coordinate within each ring. Writing the weights as 
$w_{c,r,\phi}$, this projection is given by
\begin{align}
    (\Pi_\text{ring} W)_{c,r,\phi}=\frac{1}{N_\phi}\sum_{\phi'=0}^{N_{\phi}-1}w_{c,r,\phi'}.
\end{align}
We define the following ring-invariance score
\begin{align}
    S_{\text{ring}}=1-\frac{\| W-\Pi_\text{ring} W\|_F}{\|W\|_F},\label{ring_score}
\end{align}
which measures how close the classifier is to depending only on ring-averaged features. Here, $\|\cdot\|_F$ is the Frobenius norm. By construction, $0\le S_{\text{ring}}\le 1$ and $S_{\text{ring}}=1$ corresponds to perfect constancy along $\phi$ within each ring. The results for all datasets are shown in Table~\ref{tab:ring_score}. The STM dataset yields the highest score, followed by RotFMNIST and RotMNIST, whereas MNIST and CIFAR achieve substantially lower values. This is consistent with the use of rotational augmentation for STM, RotFMNIST and RotMNIST, which naturally biases the classifier toward learning weight patterns that are more closely aligned with rotational invariance.

\section{Additional numerical results}

\begin{figure*}[t]
    \centering
    \includegraphics[width=\linewidth]{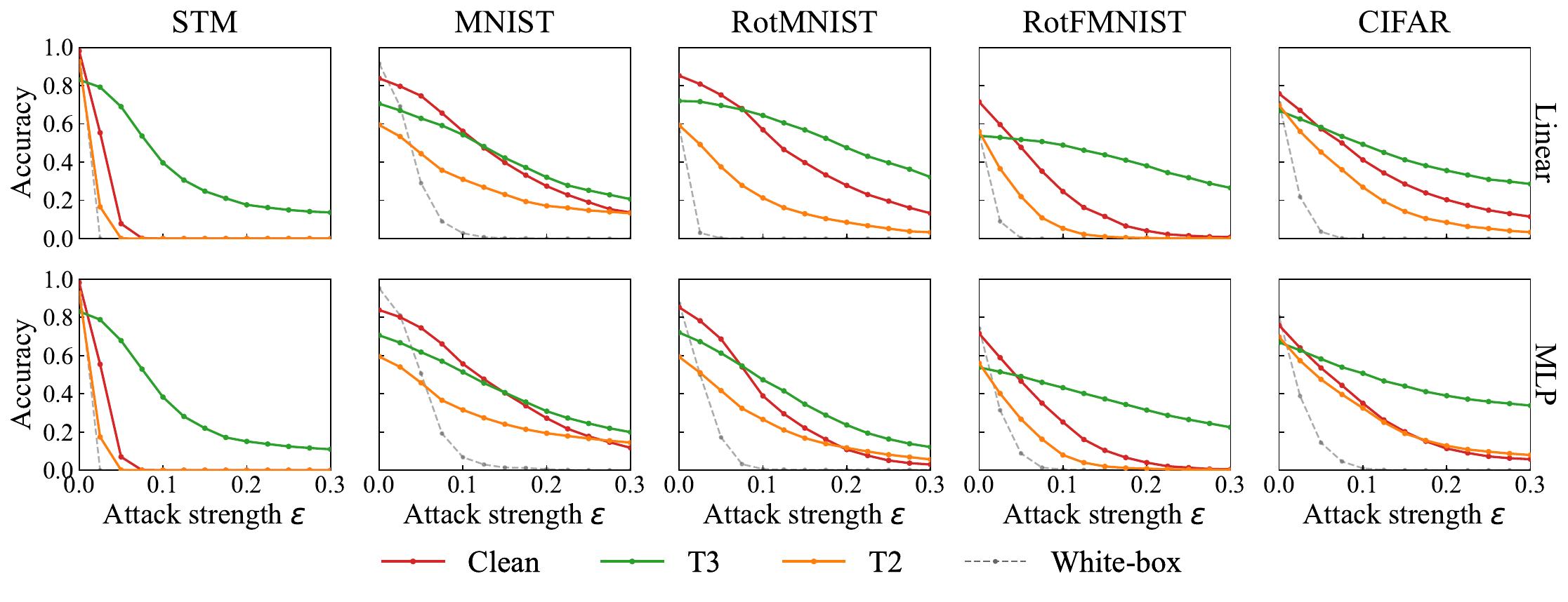}
    \caption{
    Transfer-attack test accuracy of the rotationally equivariant quantum model across the STM, MNIST, RotMNIST, RotFMNIST, and CIFAR datasets under different training-data transformations. Rows 1--2 correspond to transfer attacks generated by clean-trained LC and MLP surrogate models, respectively. Adversarial examples are generated using FGSM on 500 test examples with $\ell_{\infty}$ perturbation strength $\varepsilon$. The gray dashed curve indicates the corresponding surrogate white-box baseline, i.e., the surrogate accuracy on its own adversarial examples. Colored curves show the mean test accuracy of the target model across five runs: T3-trained (green), T2-trained (orange), and the clean-trained baseline (red).
    }
    \label{fig:adv_robust_fgsm}
\end{figure*}

For completeness, Fig.~\ref{fig:adv_robust_fgsm} reports the FGSM version of the analysis in Sec.~\ref{sub:rob_features} for the robustness evaluation of the rotationally equivariant model under transfer attacks generated by the LC and MLP surrogates. The observed trends are qualitatively consistent with the PGD-based results discussed in Sec.~\ref{sub:rob_features}. Note that for a linear classifier under the standard $\ell_\infty$ threat model, FGSM is generally not expected to be substantially weaker than PGD~\cite{goodfellow2015}.

\begin{figure*}[t]
    \centering
    \includegraphics[width=\linewidth]{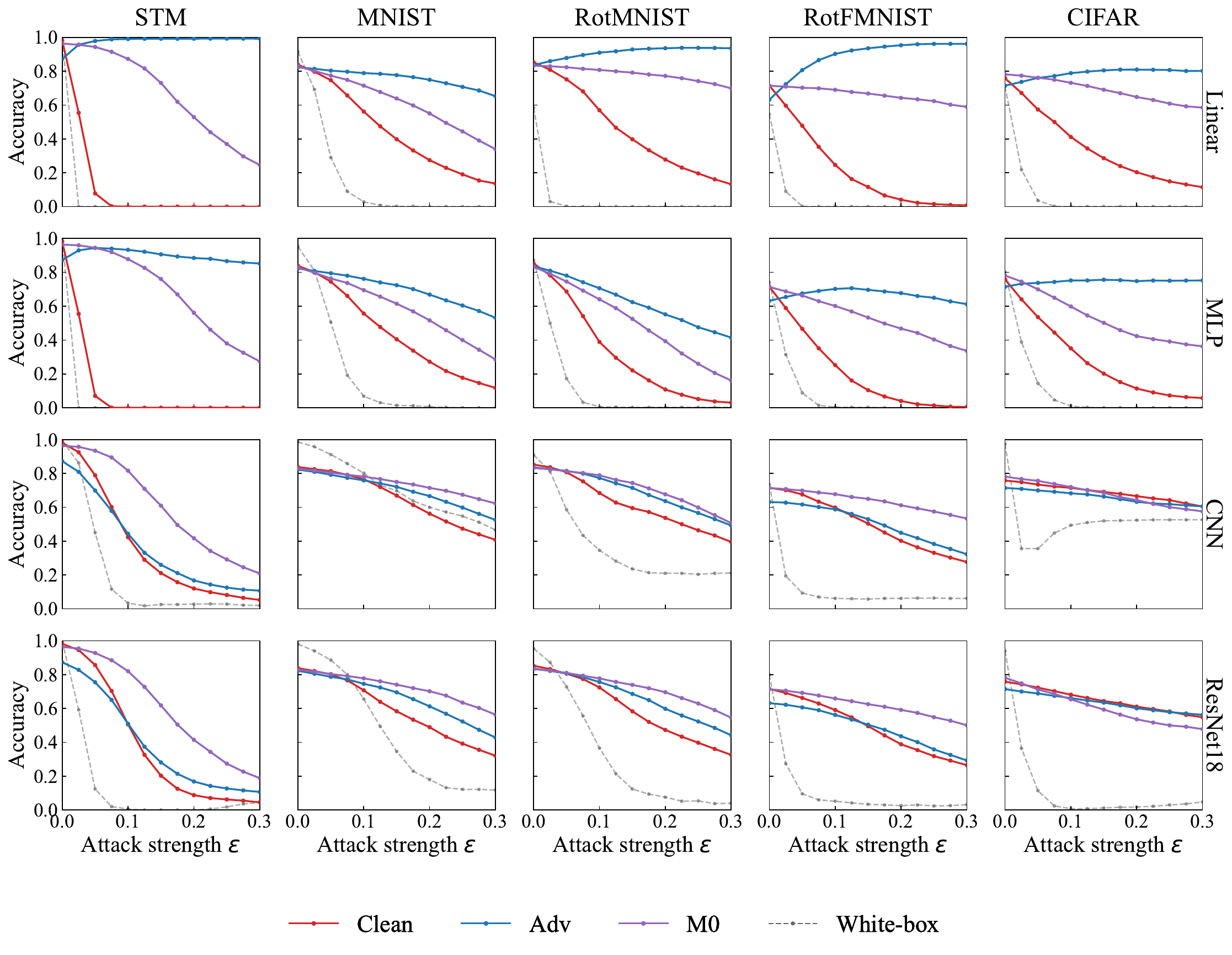}
    \caption{
    Transfer-attack test accuracy of the rotationally equivariant quantum model across the STM, MNIST, RotMNIST, RotFMNIST, and CIFAR datasets under adversarial training and readout modification. Rows 1--4 correspond to transfer attacks generated by clean-trained LC, MLP, CNN, and ResNet18 surrogate models, respectively. Adversarial examples are generated using FGSM on 500 test examples with $\ell_{\infty}$ perturbation strength $\varepsilon$. The gray dashed curve indicates the corresponding surrogate white-box baseline, i.e., the surrogate accuracy on its own adversarial examples. Colored curves show the mean test accuracy of the target model across five runs: adversarially trained (blue), clean-trained with the $m=0$ mode projected out at measurement (purple), and the clean-trained baseline (red).
    }
    \label{fig:adv_training_fgsm}
\end{figure*}

Fig.~\ref{fig:adv_training_fgsm} reports the corresponding FGSM-based version of the robustness evaluation in Sec.~\ref{sub:adv_rob}. Again, the observed trends are qualitatively consistent with the PGD-based results discussed in Sec.~\ref{sub:adv_rob}.

\end{document}